\documentclass[12pt,english,reprint,amsmaths,amssymb,prl]{revtex4-1}
\usepackage[T1]{fontenc}
\usepackage[latin9]{inputenc}
\usepackage{color}
\usepackage{babel}
\usepackage{array}
\usepackage{refstyle}
\usepackage{booktabs}
\usepackage{units}
\usepackage{textcomp}
\usepackage{amstext}
\usepackage{graphicx}
\usepackage[unicode=true,pdfusetitle,
 bookmarks=false,
 breaklinks=false,pdfborder={0 0 0},pdfborderstyle={},backref=false,colorlinks=true]
 {hyperref}

\makeatletter

\AtBeginDocument{\providecommand\secref[1]{\ref{sec:#1}}}
\AtBeginDocument{\providecommand\tabref[1]{\ref{tab:#1}}}
\AtBeginDocument{\providecommand\figref[1]{\ref{fig:#1}}}
\AtBeginDocument{\providecommand\subsecref[1]{\ref{subsec:#1}}}
\AtBeginDocument{\providecommand\eqref[1]{\ref{eq:#1}}}
\providecommand{\tabularnewline}{\\}
\newcommand{\lyxdot}{.}

\RS@ifundefined{subsecref}
  {\newref{subsec}{name = \RSsectxt}}
  {}
\RS@ifundefined{thmref}
  {\def\RSthmtxt{theorem~}\newref{thm}{name = \RSthmtxt}}
  {}
\RS@ifundefined{lemref}
  {\def\RSlemtxt{lemma~}\newref{lem}{name = \RSlemtxt}}
  {}

\renewcommand{\tabref}{\Tabref}
\renewcommand{\figref}{\Figref}
\renewcommand{\secref}{\Secref}
\renewcommand{\subsecref}{\Subsecref}
\usepackage[shortcuts]{extdash}

\makeatother

\begin{document}

\title{Causes of ferroelectricity in HfO$_{2}$-based thin films: An \emph{ab
initio} perspective}

\author{Mehmet Dogan{*}$^{1,2,3,4}$, Nanbo Gong$^{1,5}$, Tso-Ping Ma$^{1,5}$
and Sohrab Ismail-Beigi$^{1,2,6,7}$}

\affiliation{$^{1}$Center for Research on Interface Structures and Phenomena,
Yale University, New Haven, Connecticut 06520, USA $\linebreak$ $^{2}$Department
of Physics, Yale University, New Haven, Connecticut 06520, USA $\linebreak$
$^{3}$Department of Physics, University of California, Berkeley,
California 94720, USA $\linebreak$ $^{4}$Materials Science Division,
Lawrence Berkeley National Laboratory, Berkeley, California 94720,
USA $\linebreak$ $^{5}$Department of Electrical Engineering, Yale
University, New Haven, Connecticut 06520, USA $\linebreak$ $^{6}$Department
of Applied Physics, Yale University, New Haven, Connecticut 06520,
USA $\linebreak$ $^{7}$Department of Mechanical Engineering and
Materials Science, Yale University, New Haven, Connecticut 06520,
USA $\linebreak$ {*}Corresponding author: mhmtdogan@gmail.com}

\date{April 1, 2019}
\begin{abstract}
We present a comprehensive first principles study of doped hafnia
in order to understand the formation of the ferroelectric $\text{orthorhombic}\left[001\right]$
grains. Assuming that tetragonal grains are present during the early
stages of growth, matching plane analysis shows that $\text{tetragonal}\left[100\right]$
grains can transform into $\text{orthorhombic}\left[001\right]$ during
thermal annealing, when they are laterally confined by other grains.
We show that among 0\%, 2\% and \%4 Si doping, 4\% doping provides
the best conditions for the $\text{tetragonal}\left[100\right]\rightarrow\text{orthorhombic}\left[001\right]$
transformation. This also holds for Al doping. We also show that for
Hf$_{x}$Zr$_{1-x}$O$_{2}$, where we have studied $x=1.00,0.75,0.50,0.25,0.00,$
the value $x=0.50$ provides the most favorable conditions for the
desired transformation. In order for this transformation to be preferred
over the $\text{tetragonal}\left[100\right]\rightarrow\text{monoclinic}\left[100\right]$
transformation, out-of-plane confinement also needs to be present,
as supplied by a top electrode. Our findings illuminate the mechanism
that causes ferroelectricity in hafnia-based films and provide an
explanation for common experimental observations for the optimal ranges
of doping in Si:HfO$_{2}$, Al:HfO$_{2}$ and Hf$_{x}$Zr$_{1-x}$O$_{2}$.
We also present model thin film heterostructure computations of Ir/HfO$_{2}$/Ir
stacks in order to isolate the interface effects, which we show to
be significant.
\end{abstract}
\maketitle

\section{Introduction\label{sec:Introduction6}}

Achieving ferroelectricity in thin films has been a decades-long research
endeavor because of potential technological applications, e.g., the
ferroelectric field effect transistors (FEFET) and ferroelectric random-access
memory (FERAM) \citep{mckee2001physical,hoffman2010ferroelectric,kumah2016engineered,dogan2018singleatomic}.
The ferroelectric-based technology provides tremendous advantages
over the dominant non-volatile memory technology, such as low power
consumption, controllability over variability, and fast switching
speed. However, this promising technology has not been widely implemented
due to the lack of ferroelectric materials that fulfill the all of
the requirements for a viable memory technology: scalability, CMOS
compatibility, and memory retention \citep{ma2002whyis,dunkel2017afefet}.
HfO$_{2}$ is one of the most widely used dielectric gate oxides in
today's field effect transistor devices \citep{gong2016whyis}. Most
importantly, it is thermodynamically stable on silicon up to high
temperatures, allowing abrupt HfO$_{2}$/Si interfaces to be grown
without formation of silica in the interfacial region \citep{chiou2007characteristics}.
In addition, with a large band gap of 5.3 - 5.7 eV and a dielectric
constant of $\varepsilon_{r}\approx20$ in its bulk form under ordinary
conditions \citep{park2015ferroelectricity}, HfO$_{2}$ is a widely
used gate insulator and a replacement material for SiO$_{2}$ \citep{wilk2001highkappa}.
The recent discovery of ferroelectricity in HfO$_{2}$-based thin
films has further multiplied the research interest in this material
\citep{boscke2011ferroelectricity,muller2012ferroelectricity}. It
has been shown in various experimental studies that ferroelectricity
in HfO$_{2}$-based films arises from the creation of the polar orthorhombic
phase (space group: Pca$2_{1}$) of HfO$_{2}$ during a rapid annealing
process in conjunction with the presence of a capping electrode (typically
TiN). It has also been demonstrated that the ferroelectric properties
of these films strongly depend on factors such as the doping species,
doping concentration, annealing temperature and film thickness \citep{park2015ferroelectricity,gong2018nucleation}.
Even though HfO$_{2}$-based ferroelectric memory have been experimentally
demonstrated using various conditions, a systematic microscopic understanding
of the effects of the aforementioned factors is presently lacking.
This is, in part, due to the polycrystalline and complex nature of
the HfO$_{2}$ films that have been grown to date, and in part, due
to the relatively new interest in this field.

To the best of our knowledge, the ferroelectric hafnia-based thin
films are polycrystalline and contain differently oriented grains
of monoclinic (space group: P$2_{1}$/s), tetragonal (P$4_{2}$/nmc)
and orthorhombic (Pca$2_{1}$) phases in various ratios. The monoclinic
and the tetragonal phases are non-polar, and they are the observed
bulk phases of HfO$_{2}$ at room temperature and at high temperature,
respectively \citep{wang1992hafniaand,zhao2002firstprinciples}. Experiments
demonstrate that the orthorhombic phase arises during the rapid annealing
with a capping electrode. In addition, the concentration of dopants
is crucial in determining the ferroelectric properties. Because the
volume fraction of the orthorhombic phase compared to the other non-polar
phases is what ultimately decides the robustness of ferroelectricity
in the HfO$_{2}$ films, a structural understanding of the favorable
conditions for this phase is crucial in order to optimize the growth
procedure.

To this end, in our \emph{ab initio} studies we investigate the energetics
of different phases of HfO$_{2}$ with varying amounts of Si and Zr
doping and subject to a range of epitaxial strain states. In \secref{Experiment6}
we summarize our knowledge of the experimental findings on ferroelectric
thin films of hafnia to date. After describing our methods in \ref{sec:Methods6},
we move on to our computational study of doping and strain on hafnia
with a particular focus on Si:HfO$_{2}$ and Hf$_{x}$Zr$_{1-x}$O$_{2}$
in \secref{Results6}. We have found that at certain doping levels
the transformation of the high temperature tetragonal phase to the
out-of-plane polarized orthorhombic phase is favored. These results,
together with additional analysis, help explain the common experimental
observations as well as some of the underlying causes from a microscopic
viewpoint. We also describe, in \secref{Results6}, results on simulated
HfO$_{2}$ thin films including the interface with electrodes in order
to obtain some insight on the energetics of interfacial and surface
effects in thin films.

\section{Summary of experiments to date\label{sec:Experiment6}}

Since their discovery in 2011, ferroelectric hafnia thin films have
garnered tremendous experimental attention. In \tabref{Experimental}
we list some of the studies that have investigated factors such as
doping species, doping concentration, thickness of the film, and top
and bottom electrodes (TE/BE).

\begin{table}
\begin{centering}
\begin{tabular}{|>{\centering}p{0.185\columnwidth}|>{\centering}p{0.14\columnwidth}|>{\centering}p{0.17\columnwidth}|>{\centering}p{0.12\columnwidth}|>{\centering}p{0.295\columnwidth}|}
\hline 
Ref. & Dopant & TE/BE & $d$ (nm) & Observations\tabularnewline
\hline 
\hline 
Böscke et al. \citep{boscke2011ferroelectricity} & Si & TiN/TiN & 7-10 & FE at 2.6\%-4.3\% Si; $\sim$AFE at 5.6\% Si.\tabularnewline
\hline 
Müller et al. \citep{muller2011ferroelectricity} & Y & TiN/TiN & 10 & FE at 2.3\%-5.2\% Y.\tabularnewline
\hline 
Mueller et al. \citep{mueller2012incipient} & Al & TiN/TiN & 16 & FE at 4.8\% Al; AFE at 8.5\% Al.\tabularnewline
\hline 
Müller et al. \citep{muller2012ferroelectricity} & Zr & TiN/TiN & 9 & PE at <30\% Zr; FE at \%30-\%60 Zr; AFE at >70\% Zr.\tabularnewline
\hline 
Yurchuk et al. \citep{yurchuk2013impactof} & Si & TiN/TiN & 9, 27 & FE at 4.4\% Si, 9 nm; $\sim$PE at 27 nm.\tabularnewline
\hline 
Park et al. \citep{park2013evolution} & 50\% Zr & TiN/TiN & 5-25 & FE at 5-17 nm; PE at 25 nm. Pca2$_{1}$ phase confirmed.\tabularnewline
\hline 
Park et al. \citep{park2014theeffects} & 50\% Zr & TiN/TiN, TiN/Pt & 5-27 & FE at 8-19 nm for TiN; Less FE for Pt at 8 nm; PE for Pt at > 13 nm.\tabularnewline
\hline 
Lomenzo et al. \citep{lomenzo2014ferroelectric} & Si & TiN/Si, Ir/Si, Ir/Ir & 10 & FE similar for TiN/Si and Ir/Si. Smaller $P_{r}$ for Ir/Ir.\tabularnewline
\hline 
Schroeder et al. \citep{schroeder2014impactof} & Si, Al,

Y, Gd,

La, Sr & TiN/TiN & 10 & FE at 4.4\% Si; AFE at >5.6\% Si. Similar for Al. No AFE for other
dopants.\tabularnewline
\hline 
Park et al. \citep{park2014studyon} & 50\% Zr & TiN/TiN, TiN/Ir & 9-24 & FE at 9-19 nm for TiN; FE at 12-15 nm for Ir.\tabularnewline
\hline 
Sang et al. \citep{sang2015onthe} & Gd & TiN/TiN & 27 & FE; Pca2$_{1}$ phase confirmed.\tabularnewline
\hline 
Hoffmann et al. \citep{hoffmann2015stabilizing} & Gd & TiN/TiN, TiN/TaN, TaN/TaN & 10-27 & FE similar for all stacks; TaN/TaN > TiN/TiN $\simeq$ TiN/TaN in
terms of $P_{r}$.\tabularnewline
\hline 
Chernikova et al. \citep{chernikova2016ultrathin} & 50\% Zr & TiN/TiN & 2.5 & FE; Pca2$_{1}$ phase confirmed.\tabularnewline
\hline 
Park et al. \citep{park2017effectof} & Al, Gd & TiN/TiN & 10 & FE at 5.7\%-6.9\% Al, 3.0\%-3.9\% Gd; PE at 8.8\% Al.\tabularnewline
\hline 
Park et al. \citep{park2018originof} & Si & TiN/TiN & 10, 40 & FE at 3.8\%-5.6\% Si, 10 nm; $\sim$PE at 4.5\% Si, 40 nm; PE at 5.0\%-6.3\%
Si, 40 nm.\tabularnewline
\hline 
\end{tabular}
\par\end{centering}
\caption[Selected experimental studies of ferroelectric hafnia thin films published
between 2011 and 2018, listed chronologically.]{\label{tab:Experimental}Selected experimental studies of ferroelectric
hafnia thin films published between 2011 and 2018, listed chronologically.
PE means paraelectric, FE means ferroelectric, AFE means antiferroelectric.}
\end{table}

These studies have generally found that, when the HfO$_{2}$ films
are doped with a few percent of a wide range of dopants or with $\sim$50\%
Zr, grown to be $\sim5-20$ nm thick between two metal electrodes,
and annealed at $\sim800-1000$ °C, they can display ferroelectricity.
Some of the studies \citep{muller2011ferroelectricity,boscke2011ferroelectricity,mueller2012incipient,lomenzo2014ferroelectric}
have shown that for TiN/HfO$_{2}$/TiN stacks, if the film is annealed
before the deposition of the top electrode, then the ferroelectric
behavior is significantly suppressed, which indicates that the confinement
provided by the top electrode during annealing is crucial for ferroelectricity.

In terms of the atomic structure of the films, many of these studies
have performed XRD analyses to show that ferroelectricity is intimately
related to the presence of the orthorhombic Pca2$_{1}$ phase of HfO$_{2}$.
The path to the stabilization of this phase is believed to be via
the creation of grains of the tetragonal P$4_{2}$/nmc phase which
are present when the film is first deposited \citep{muller2012ferroelectricity}.
The tetragonal phase is known to be stabilized by surface effects
\citep{garvie1978stabilization}, and recent studies have shown doping
to be a stabilizer for this phase as well \citep{tomida2006dielectric,boscke2009increasing}.
Hence, it appears that for ferroelectric hafnia-based films, the orthorhombic
phase is obtained from the tetragonal grains during rapid thermal
annealing under the confinement of a top electrode. Crystallization
without a top electrode leads to the formation of the monoclinic P$2_{1}$/s
phase, which is the non-polar ground state at relevant temperatures.
The causes for the favorability of the tetragonal $\rightarrow$ orthorhombic
transition over the tetragonal $\rightarrow$ monoclinic transition
under these doping/thickness/temperature/confinement conditions are
not well understood.

\section{Computational methods\label{sec:Methods6}}

\begin{figure}
\begin{centering}
\includegraphics[width=0.95\columnwidth]{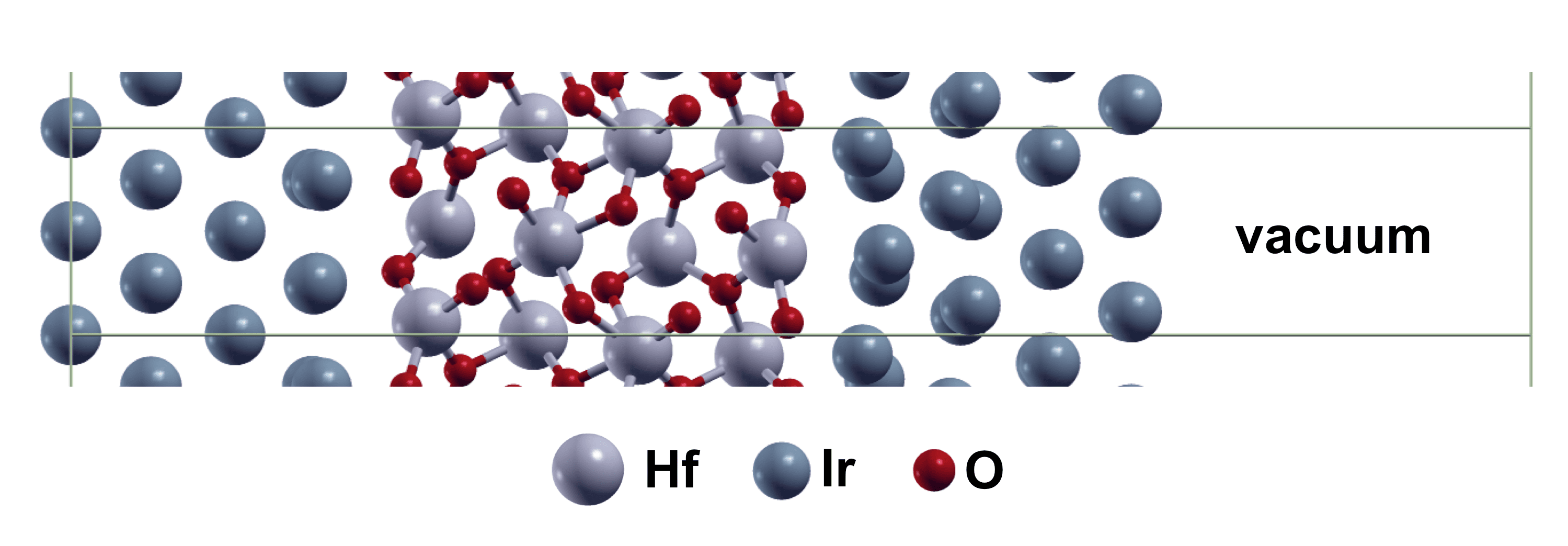}
\par\end{centering}
\caption[A sample supercell for a thin film relaxation of an Ir/HfO/Ir stack.]{\label{fig:Simcell}A sample simulation cell for a thin film relaxation
of an Ir/HfO$_{2}$/Ir stack. HfO$_{2}$ is in the monoclinic phase
with $\left[001\right]$ orientation. The in-plane lattice is fixed
to the lattice parameters of this phase and orientation of HfO$_{2}$
(see \subsecref{Thin-film} for details of these thin film simulations).
Periodic copies of the stack are separated by vacuum.}
\end{figure}

We compute minimum energy structures using density functional theory
(DFT) in the Perdew\--Burke\--Ernzerhof Generalized Gradient Approximation
(PBE GGA) \citep{perdew1996generalized} with ultrasoft pseudopotentials
\citep{vanderbilt1990softselfconsistent}. We use the QUANTUM ESPRESSO
software package \citep{giannozzi2009quantum}. Plane wave energy
cutoffs of $55$ Ry and $35$ Ry are used for bulk and thin film simulations,
respectively. We sample the Brillouin zone with an $8\times8\times8$
Monkhorst\--Pack $k$-point mesh (per 12 atom unit cells) and a $0.02$
Ry Marzari\--Vanderbilt smearing \citep{marzari1999thermal} for
bulk samples\textcolor{black}{; slab (thin film) systems are sampled
with a} $6\times6\times1$\textcolor{black}{{} mesh. For thin fi}lm
simulations, periodic copies of the slab are separated by $\sim12\text{\AA}$
of vacuum in the $z-$direction (see \figref{Simcell} for a representative
unit cell), and the in-plane lattice constants of the slab are fixed
to the computed bulk lattice constants of HfO$_{2}$ for the plane
under consideration. In general, a slab may have an overall dipole
moment that can artificially interact with its periodic copies through
the vacuum gap. We eliminate this effect by introducing a fictitious
dipole in the vacuum region of the cell which cancels out the electric
field in vacuum \citep{bengtsson1999dipolecorrection}. All atoms
are relaxed until the forces on the atoms are less than $10^{-4}\:\textrm{Ry/Bohr}$
in magnitude along all axial directions. We use both direct substitution
of atoms as well as the virtual crystal approximation (VCA) to model
doping \citep{bellaiche2000virtual}. In the VCA, to approximate the
mixing of two elements $A$ and $B$ with ratios $x$ and $1-x$,
a virtual element is created by linearly interpolating the pseudopotentials
of $A$ and $B$ such that the resulting pseudopotential is $V=xV_{A}+\left(1-x\right)V_{B}$.
The VCA is known to be a good approximation when alloying chemically
similar elements with the same valence state. A detailed comparison
of these two approaches is reported in \subsecref{Effects-of-doping}.

\section{Results \label{sec:Results6}}

\subsection{Bulk phases of HfO$_{2}$\label{subsec:Bulk-phases}}

\begin{figure}
\begin{centering}
\includegraphics[width=0.6\columnwidth]{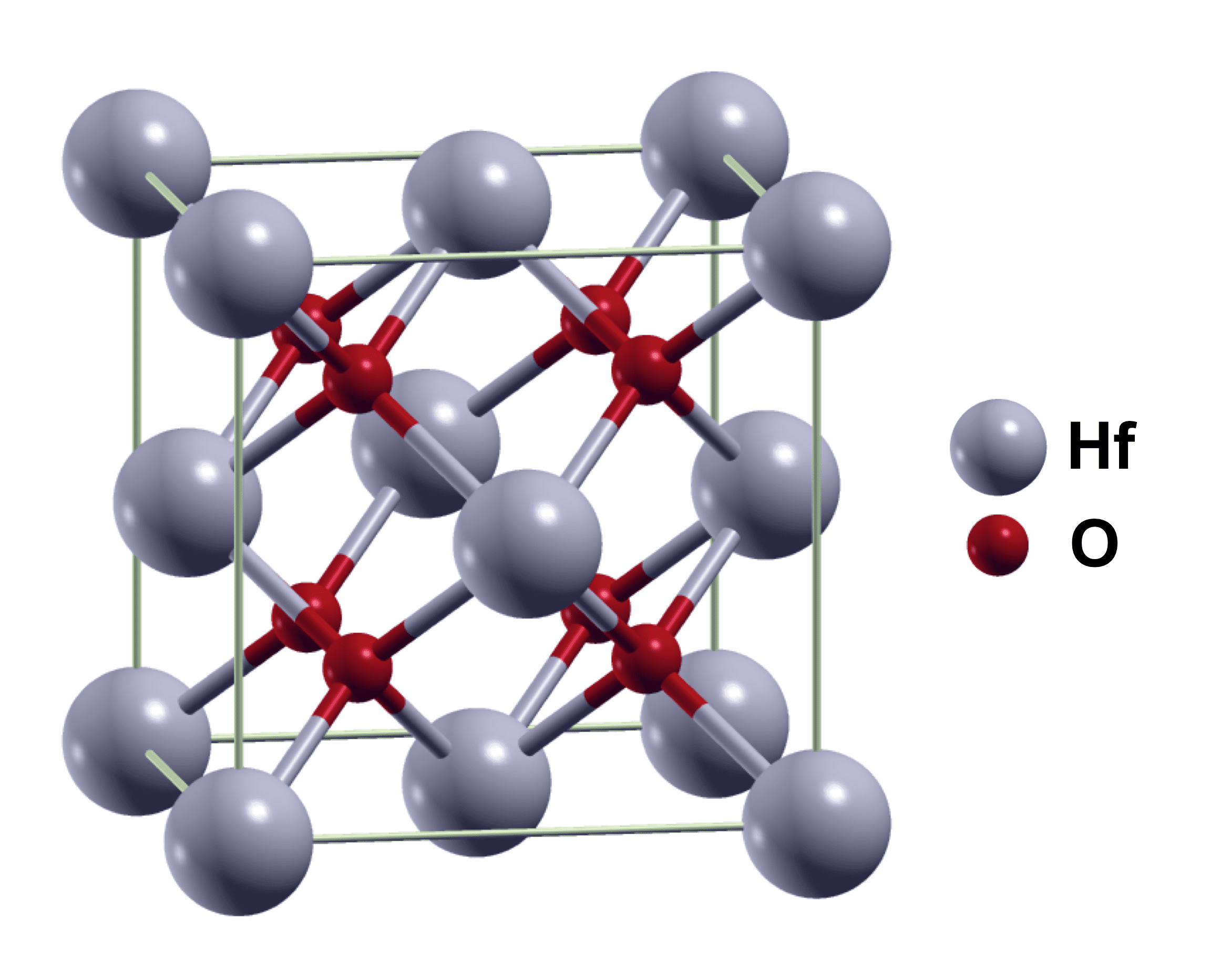}
\par\end{centering}
\caption{\label{fig:Cubic-phase}Cubic phase of HfO$_{2}$ (space group: Fm$\overline{3}$m),
where hafnium atoms occupy the fcc lattice sites.}

\end{figure}

HfO$_{2}$ can be observed in three phases in its bulk form. The monoclinic
phase (space group: P$2_{1}$/c) is stable all the way up to $\sim$2000
K. Between $\sim$2000 K and $\sim$2900 K, the tetragonal phase (space
group: P$4_{2}$/nmc) is observed. The highest symmetry cubic phase
(space group: Fm$\overline{3}$m) is observed between $\sim$2900
K and the melting temperature of $\sim$3100 K \citep{wang1992hafniaand,zhao2002firstprinciples}.
The cubic phase of HfO$_{2}$ is depicted in \figref{Cubic-phase}.
It is a face centered cubic structure with one formula unit per lattice
point. The tetragonal and the monoclinic phases are obtained from
the cubic phase through consecutive symmetry breaking operations.

\begin{figure}
\begin{centering}
\includegraphics[width=1\columnwidth]{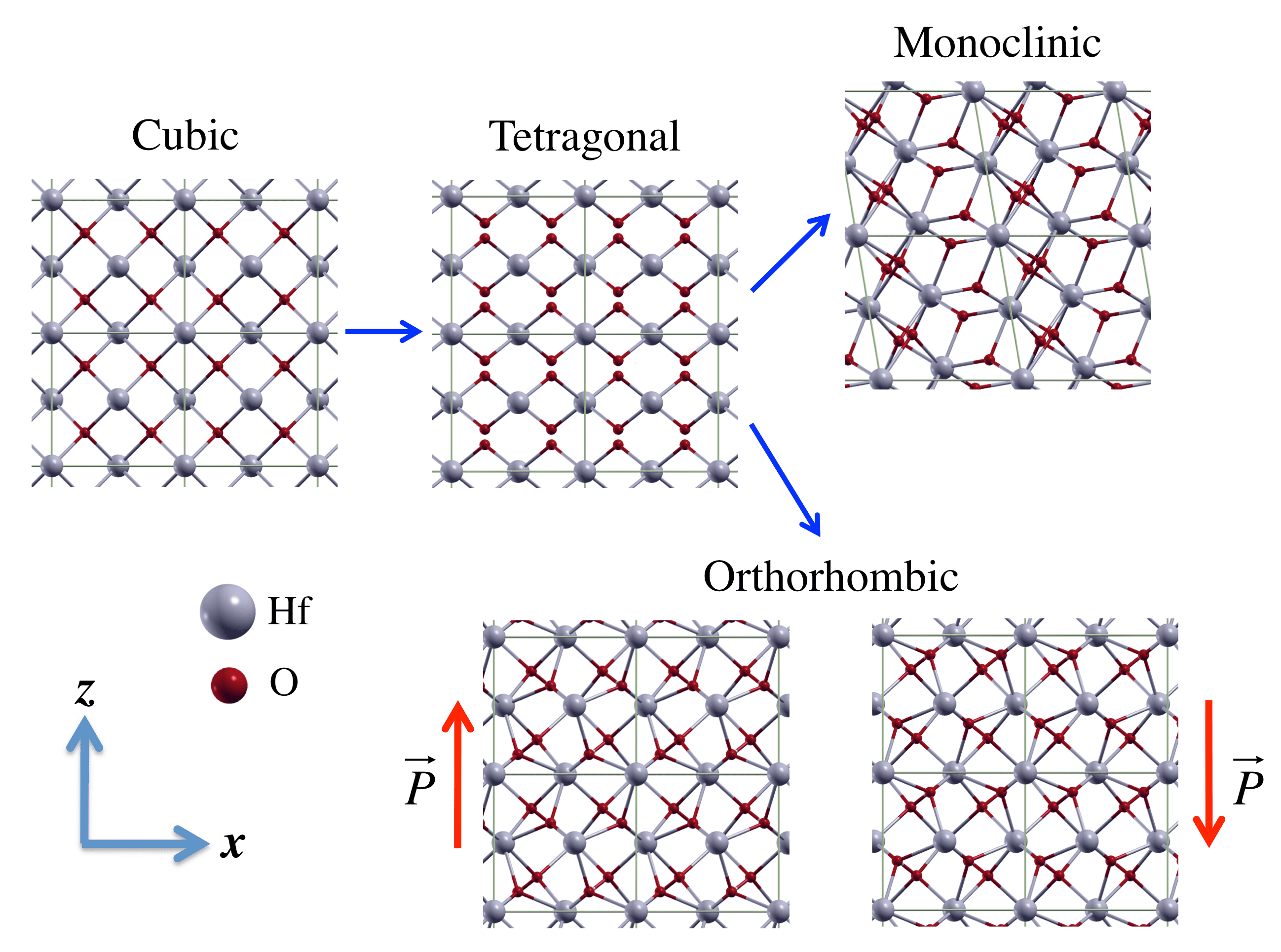}
\par\end{centering}
\caption[Bulk phases of HfO$_{2}$ considered in this work.]{\label{fig:Bulk_phases}Bulk phases of HfO$_{2}$ considered in this
work. Subgroup relations are shown by blue arrows. The cubic, tetragonal
and monoclinic phases are the experimentally observed bulk phases.
The non-centrosymmetric orthorhombic phase is observed in some thin
films of HfO$_{2}$ and gives rise to ferroelectricity. For each phase,
the 12-atom simulation cell that we use in this study is shown by
thin straight lines.}
\end{figure}

These three bulk phases are all centrosymmetric, causing the bulk
oxide to be paraelectric. However, as described in \secref{Experiment6},
the recent discovery of ferroelectricity in HfO$_{2}$ thin films
indicates that a non-centrosymmetric orthorhombic phase (space group:
Pca$2_{1}$) is stabilized under certain growth conditions that gives
rise to a switchable polarization. The four phases of HfO$_{2}$ that
are the focus of this study are shown in \figref{Bulk_phases}. The
orthorhombic and monoclinic phases are obtained from the tetragonal
phase by symmetry breaking operations, indicated by a blue arrow in
the figure.

In \tabref{Relative-energies} we compare our computed energies of
each phase (relative to the monoclinic phase) with the results from
previous computational studies. We find that our results agree well
with Ref. \citep{zeng2014evolutionary} which uses the GGA to approximate
the exchange-correlation functional; and Refs. \citep{reyes-lillo2014antiferroelectricity}
and \citep{materlik2015theorigin} which use the Local Density Approximation
(LDA) agree with each other.

\begin{table}
\begin{centering}
\begin{tabular}{>{\centering}p{0.24\columnwidth}>{\centering}p{0.17\columnwidth}>{\centering}p{0.17\columnwidth}>{\centering}p{0.17\columnwidth}>{\centering}p{0.17\columnwidth}}
\hline 
\noalign{\vskip0.3cm}
Bulk phase & This work (GGA) & Ref. \citep{zeng2014evolutionary} (GGA) & Ref. \citep{reyes-lillo2014antiferroelectricity} (LDA) & Ref. \citep{materlik2015theorigin} (LDA)\tabularnewline[0.3cm]
\hline 
\noalign{\vskip0.1cm}
\hline 
\noalign{\vskip0.3cm}
mono P$2_{1}$/c & $\equiv0.00$ & $\equiv0.00$ & $\equiv0.00$ & $\equiv0.00$\tabularnewline[0.3cm]
\hline 
\noalign{\vskip0.3cm}
ortho Pca$2_{1}$ & $0.08$ & $0.06$ & $0.05$ & $0.06$\tabularnewline[0.3cm]
\hline 
\noalign{\vskip0.3cm}
tetra P$4_{2}$/nmc & $0.16$ & $0.14$ & $0.07$ & $0.09$\tabularnewline[0.3cm]
\hline 
\noalign{\vskip0.3cm}
cubic Fm$\overline{3}$m & $0.27$ & $0.21$ & $0.09$ & $0.14$\tabularnewline[0.3cm]
\hline 
\end{tabular}
\par\end{centering}
\caption{\label{tab:Relative-energies}Energies of the bulk phases of HfO$_{2}$
in eV per HfO$_{2}$ (relative to the monoclinic phase) considered
in this work, compared with previous computational work.}

\end{table}

In \tabref{Lat_para} we compare lattice parameters with the results
of previous studies. We find that our results generally lie within
the range of agreement among the previous works where there is a range
of <5\% for a given lattice parameter.

\begin{table*}
\begin{centering}
\begin{tabular}{>{\centering}p{0.13\textwidth}>{\centering}m{0.1\textwidth}>{\centering}m{0.17\textwidth}>{\centering}m{0.17\textwidth}>{\centering}m{0.17\textwidth}>{\centering}m{0.17\textwidth}}
\toprule 
\addlinespace[0.3cm]
Bulk phase & Parameters & This work ($\text{\AA}$) & Ref. \citep{zeng2014evolutionary} & Ref. \citep{huan2014pathways} & Ref. \citep{materlik2015theorigin}\tabularnewline\addlinespace[0.3cm]
\midrule
\addlinespace[0.1cm]
\midrule 
Monoclinic & $a,b,c,$

$\beta$ & 5.12, 5.18, 5.30,

$99.6^{\circ}$ & 5.09, 5.16, 5.26,

$99.7^{\circ}$ & 5.14, 5.20, 5.31,

$99.8^{\circ}$ & 5.11, 5.18, 5.29,

$99.7^{\circ}$\tabularnewline
\midrule 
Orthorhombic & $a,b,c$ & 5.25, 5.04, 5.07 & 5.11, 4.90, 4.92 & 5.29, 5.01, 5.08 & 5.23, 5.04, 5.06\tabularnewline
\midrule 
Tetragonal & $a,c$ & 5.07, 5.19 & 5.03, 5.15 & 5.06, 5.28 & 5.05, 5.14\tabularnewline
\midrule 
Cubic & $a$ & 5.06 & 5.03 & - & 5.04\tabularnewline
\bottomrule
\end{tabular}
\par\end{centering}
\caption[Lattice parameters of the bulk phases of HfO$_{2}$ compared with
previous computational studies. ]{\label{tab:Lat_para}Lattice parameters (in $\text{\AA}$) of the
bulk phases of HfO$_{2}$ compared with previous computational studies.
For the monoclinic phase, $\beta$ is the angle between $\vec{a}$
and $\vec{c}$, which is the only non-perpendicular angle for this
phase. }
\end{table*}

\subsection{Effects of doping on bulk HfO$_{2}$\label{subsec:Effects-of-doping}}

Due to its importance in the ferroelectric properties of hafnia thin
films, we have investigated the role of doping in stabilizing the
various phases of HfO$_{2}$ with respect to each other. We list the
energies of bulk phases with respect to the monoclinic phase for various
dopants in \tabref{Doped-energies}. These simulations are done with
$2\times2\times2$ simulation cells with 96 atoms, where one Hf per
cell (i.e., 1 in 32) is replaced by the dopant. All atomic positions
and cell parameters are then relaxed. This leads to a doping ratio
of 3.125\% where the dopants are equally spaced in the three lattice
directions. For elements with a different number of valence electrons
than Hf, such as N, Al, Sr, Y and La, we have additionally computed
relaxed energies with compensating electrons or holes and compared
these with the neutral relaxations. 

\begin{table}
\begin{centering}
\begin{tabular}{cccc}
\hline 
\noalign{\vskip0.3cm}
$\ $Bulk phase$\ $ & $\ $ortho Pca$2_{1}$$\ $ & $\ $tetra P$4_{2}$/nmc$\ $ & $\ $cubic Fm$\overline{3}$m$\ $\tabularnewline[0.3cm]
\hline 
\noalign{\vskip0.1cm}
\hline 
\noalign{\vskip0.2cm}
No doping & 0.09 & 0.16 & 0.26\tabularnewline[0.2cm]
\hline 
\noalign{\vskip0.2cm}
C & 0.07 & 0.08 & 0.46\tabularnewline[0.2cm]
\hline 
\noalign{\vskip0.2cm}
N & 0.10 & 0.15 & 0.41\tabularnewline[0.2cm]
\hline 
\noalign{\vskip0.2cm}
N{*} (-1e) & 0.09 & 0.10 & 0.41\tabularnewline[0.2cm]
\hline 
\noalign{\vskip0.2cm}
Al & 0.09 & 0.16 & 0.28\tabularnewline[0.2cm]
\hline 
\noalign{\vskip0.2cm}
Al{*} (+1e) & 0.09 & 0.15 & 0.33\tabularnewline[0.2cm]
\hline 
\noalign{\vskip0.2cm}
Si & 0.07 & 0.11 & 0.33\tabularnewline[0.2cm]
\hline 
\noalign{\vskip0.2cm}
Ti & 0.09 & 0.15 & 0.27\tabularnewline[0.2cm]
\hline 
\noalign{\vskip0.3cm}
Ge & 0.09 & 0.12 & 0.30\tabularnewline[0.3cm]
\hline 
\noalign{\vskip0.3cm}
Sr & 0.08 & 0.16 & 0.23\tabularnewline[0.3cm]
\hline 
\noalign{\vskip0.3cm}
Sr{*} (+2e) & 0.10 & 0.18 & 0.34\tabularnewline[0.3cm]
\hline 
\noalign{\vskip0.3cm}
Y & 0.08 & 0.17 & 0.24\tabularnewline[0.3cm]
\hline 
\noalign{\vskip0.3cm}
Y{*} (+1e) & 0.09 & 0.17 & 0.29\tabularnewline[0.3cm]
\hline 
\noalign{\vskip0.3cm}
La & 0.08 & 0.16 & 0.24\tabularnewline[0.3cm]
\hline 
\noalign{\vskip0.3cm}
La{*} (+1e) & 0.08 & 0.16 & 0.29\tabularnewline[0.3cm]
\hline 
\end{tabular}
\par\end{centering}
\caption[Energies of the bulk phases of HfO$_{2}$ in eV per HfO$_{2}$ (relative
to the monoclinic P2$_{1}$/c phase) with 3.125\% doping with various
elements.]{\label{tab:Doped-energies}Energies of the bulk phases of HfO$_{2}$
in eV per HfO$_{2}$ (relative to the monoclinic P2$_{1}$/c phase)
with 3.125\% doping with various elements. The elements with a star
({*}) denote simulations with added electrons or holes to compensate
for the difference in the number of valence electrons of that element
and Hf. The number of extra electrons per simulation cell is shown
next to each element in parentheses. For the case of N, one electron
is taken out of the system, which is denoted as (-1e).}
\end{table}

We find that doping does not change the energy difference between
the orthorhombic and the monoclinic phases significantly. In some
cases (C, N{*}, Si and Ge), it reduces the energy of the tetragonal
phase, while in some cases (C, N, N{*}, Al{*}, Si, Ge and Sr{*}) it
increases the energy of the cubic phase. We also find that changing
the number of electrons in the cell does not significantly modify
the energies of the orthorhombic and the tetragonal phases (with the
exception of N), but increases the energy of the cubic phase (again
with the exception of N). Our results are in close agreement with
the available first-principles study of doped HfO$_{2}$ that included
Ge, Sr, Y and La \citep{batra2017dopants}.

The most significant observation of this survey of dopants is that,
for non-metal dopants (C, N{*}, Si and Ge), the tetragonal phase experiences
a significant stabilization. If we focus on the monoclinic, orthorhombic
and tetragonal phases, which are the phases that participate in the
thin film processes, we conclude that, apart from the cases of C,
N{*}, Si and Ge, no significant change occurs in terms of pure phase
energetics. With these four special dopants, the reduction of energy
in the tetragonal phase relative to the monoclinic phase may favor
the formation of the tetragonal phase in thin films, and subsequently
the formation of the ferroelectric orthorhombic phase during thermal
annealing, as we will explain in \subsecref{Effects-of-strain}. To
the best of our knowledge, C and N have not been used as dopants in
HfO$_{2}$; thus it is not possible to refer to relevant experiments.
Note that C and N have atomic radii much smaller than Hf, and hence
are likely challenging to be incorporated as dopants. Ge and Ti have
been reported as dopants in HfO$_{2}$ \citep{li2005electrical,wang2012highlyuniform};
however these reports were prior to the discovery of ferroelectricity
in hafnia-based films. The other elements we have included in our
survey (Al, Sr, Y and La) have all been experimentally shown to promote
ferroelectricity in hafnia \citep{schroeder2014impactof,starschich2017anextensive,park2017effectof}.
Therefore, reduction of the energy of the tetragonal phase with respect
to the monoclinic phase, by itself, does not predict the promotion
of ferroelectricity in these films, and, it needs to be considered
in conjunction with other factors such as strain and thin film effects.

\subsubsection{Doping by Si}

\begin{figure}
\begin{centering}
\includegraphics[width=1\columnwidth]{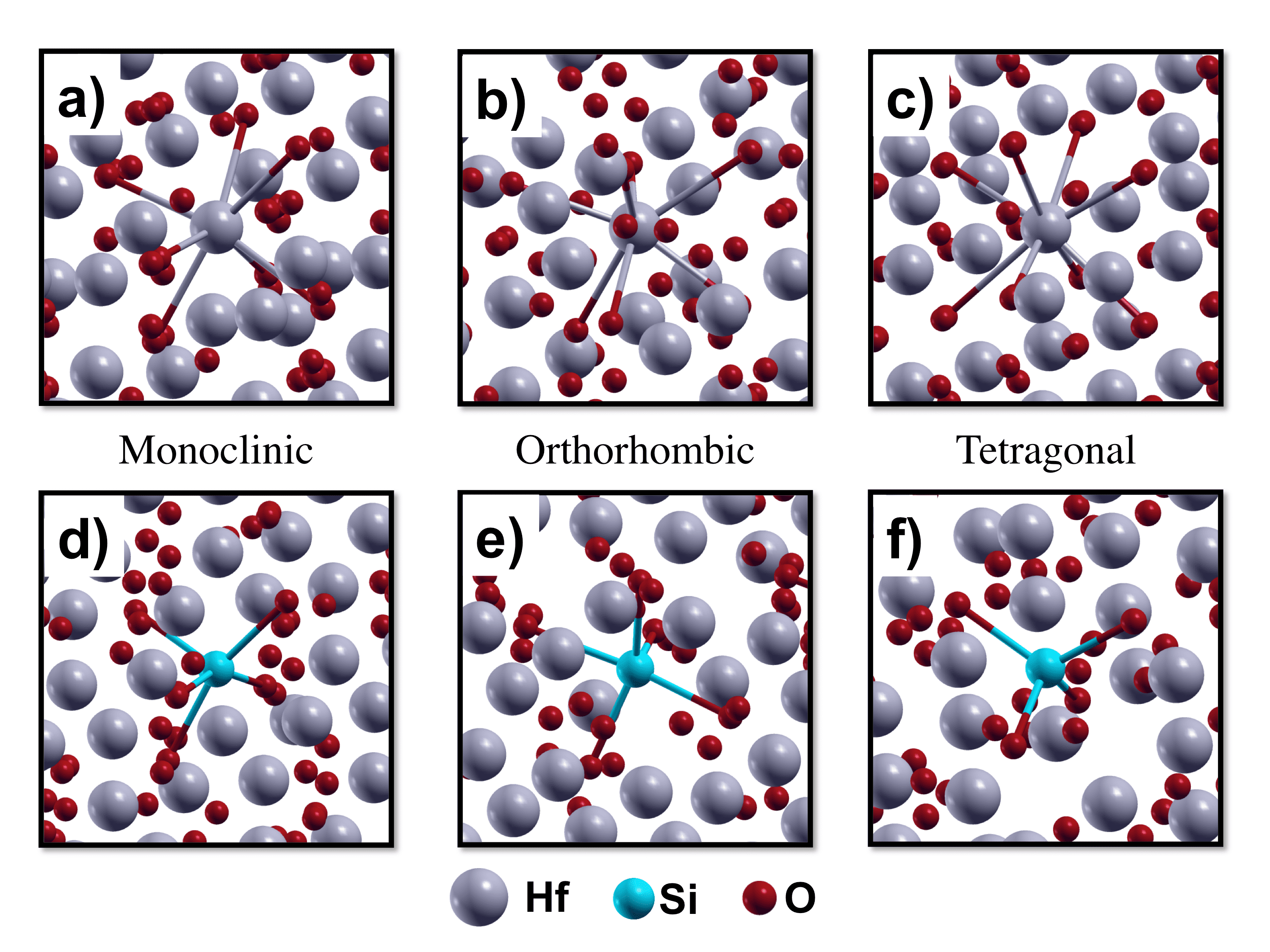}
\par\end{centering}
\caption[The environment of a Hf atom in the monoclinic, orthorhombic, and
tetragonal phases of HfO$_{2}$ compared with the environment of a
Si dopant in the same phases.]{\label{fig:Si_doping}The environment of a Hf atom in the (a) monoclinic,
(b) orthorhombic, and (c) tetragonal phases of HfO$_{2}$ compared
with the environment of a substitutional Si dopant in the (d) monoclinic,
(e) orthorhombic and (f) tetragonal phases. For each case, the bonds
between the atom in question and its nearest oxygen neighbors are
drawn. See \tabref{CN_oxygen} for the list of (Hf,Si)-O bond lengths
in each case and the number of nearest oxygen neighbors (coordination
number).}
\end{figure}

In order to gain a comprehensive understanding of the effects of doping
by one of the elements, we focus on Si which is one of the most widely
used dopants in hafnia-based thin films (along with Zr). In \figref{Si_doping},
we present the environment of a hafnium atom by showing its bonds
with the neighboring oxygen atoms for the (a) monoclinic, (b) orthorhombic,
and (c) tetragonal phases of HfO$_{2}$. We also present the environment
of a silicon atom that replaces a hafnium atom after a full relaxation,
for the the (d) monoclinic, (e) orthorhombic, and (f) tetragonal phases.

As seen by the number of Hf-O bonds drawn in \figref{Si_doping},
a hafnium atom is seven-fold coordinated by oxygen atoms in the monoclinic
and orthorhombic phases and eight-fold coordinated in the tetragonal
phase. A silicon dopant becomes five-fold coordinated in the monoclinic
and orthorhombic phases and four-fold coordinated in the tetragonal
phase. We list the (Hf,Si)-O distances in \tabref{CN_oxygen}. We
assume that if the distance between the two atoms is not much larger
than the sum of their atomic radii, the two atoms are coordinated.
For Hf-O coordination, this sum is $2.2\ \mathring{\text{A}}$, and
for Si-O coordination, it is $1.7\ \mathring{\text{A}}$, yielding
the coordination numbers in \tabref{CN_oxygen}.

\begin{table}
\begin{centering}
\begin{tabular}{|c|c|c|c|c|c|c|c|c|c|}
\hline 
\noalign{\vskip0.2cm}
Phase & \multicolumn{8}{c|}{Nearest O neighbor distances ($\mathring{\text{A}}$)} & C. N.\tabularnewline[0.2cm]
\hline 
\noalign{\vskip0.2cm}
mono Hf & 2.05 & 2.14 & 2.14 & 2.16 & 2.18 & 2.28 & 2.30 &  & 7\tabularnewline[0.2cm]
\hline 
\noalign{\vskip0.2cm}
ortho Hf & 2.04 & 2.12 & 2.13 & 2.14 & 2.14 & 2.24 & 2.27 &  & 7\tabularnewline[0.2cm]
\hline 
\noalign{\vskip0.2cm}
tetra Hf & 2.07 & 2.07 & 2.07 & 2.07 & 2.39 & 2.39 & 2.39 & 2.39 & 8\tabularnewline[0.2cm]
\hline 
\noalign{\vskip0.2cm}
mono Si & 1.76 & 1.79 & 1.80 & 1.83 & 1.87 & 2.36 & 2.92 &  & 5\tabularnewline[0.2cm]
\hline 
\noalign{\vskip0.2cm}
ortho Si & 1.75 & 1.77 & 1.81 & 1.81 & 1.92 & 2.27 & 3.07 &  & 5\tabularnewline[0.2cm]
\hline 
\noalign{\vskip0.2cm}
tetra Si & 1.69 & 1.69 & 1.69 & 1.69 & 2.75 & 2.75 & 2.75 & 2.75 & 4\tabularnewline[0.2cm]
\hline 
\end{tabular}
\par\end{centering}
\caption[List of (Hf,Si)-O bond lengths for each of the monoclinic, orthorhombic
and tetragonal phases, for the undoped case and the 3.125\% Si doped
case.]{\label{tab:CN_oxygen}List of (Hf,Si)-O bond lengths for each of
the monoclinic, orthorhombic and tetragonal phases, for the undoped
case (top three rows) and the 3.125\% Si doped case. The number of
oxygen neighbors to Hf or Si (coordination number) is reported in
the rightmost column.}

\end{table}

We find that the monoclinic and the orthorhombic phases have the same
coordination configuration in HfO$_{2}$, and the coordination of
Si dopant is approximately the same for these two phases. Hence the
energy difference between the orthorhombic and the monoclinic phases
does not significantly change upon doping. However, in the tetragonal
phase, the hafnium atom is coordinated by 8 oxygens. A closer inspection
reveals that Hf is surrounded by two concentric oxygen tetrahedra
with Hf-O distances of 2.07 $\mathring{\text{A}}$ and 2.39 $\mathring{\text{A}}$.
After replacement of this hafnium with a silicon atom, the closer
tetrahedron is pulled in and the farther tetrahedron is pushed out
so that the distances become 1.69 $\mathring{\text{A}}$ and 2.75
$\mathring{\text{A}}$. This oxygen environment for silicon is almost
identical to its environment in the ground state of bulk SiO$_{2}$.
In the P3$_{1}$21 ($\text{\ensuremath{\alpha}}$-quartz) phase of
SiO$_{2}$, silicon atoms lie in the centers of oxygen tetrahedra
with a Si-O distance of 1.63 $\mathring{\text{A}}$ (our calculated
value). We note that among the low-energy polymorphs of SiO$_{2}$,
five out of six lowest energy structures feature tetrahedral cages
\citep{zeng2014evolutionary}. Therefore, we conclude that the favorable
tetrahedral environment of Si dopant in the tetragonal phase causes
its significant stabilization with respect to the orthorhombic and
the monoclinic phases. In the Supplementary Material, we list the
dopant-O bond lengths for all the other dopants (C, N, N{*}, Al, Al{*},
Ti, Ge, Sr, Sr{*}, Y, Y{*}, La and La{*}).

\begin{table}
\begin{centering}
\begin{tabular}{>{\centering}p{0.1\columnwidth}>{\centering}p{0.12\columnwidth}>{\centering}p{0.17\columnwidth}>{\centering}p{0.17\columnwidth}>{\centering}p{0.17\columnwidth}>{\centering}p{0.17\columnwidth}}
\toprule 
\addlinespace[0.2cm]
Phase & HfO$_{2}$ & 2\% Si (VCA) & 3.125\% Si (AS) & 3.125\% Si (VCA) & 4\% Si (VCA)\tabularnewline\addlinespace[0.2cm]
\midrule 
\addlinespace[0.2cm]
ortho & 0.08 & 0.07 & 0.07 & 0.05 & 0.03\tabularnewline\addlinespace[0.2cm]
\midrule 
\addlinespace[0.2cm]
tetra & 0.16 & 0.15 & 0.11 & 0.11 & 0.08\tabularnewline\addlinespace[0.2cm]
\midrule 
\addlinespace[0.2cm]
cubic & 0.26 & 0.17 & 0.33 & 0.12 & 0.08\tabularnewline\addlinespace[0.2cm]
\bottomrule
\end{tabular}
\par\end{centering}
\caption[Energies of the orthorhombic, tetragonal and cubic phases with respect
to the monoclinic phase, for pure and Si doped HfO$_{2}$.]{\label{tab:Si_doping}Energies of the orthorhombic, tetragonal and
cubic phases with respect to the monoclinic phase, for pure and Si
doped HfO$_{2}$, as computed by direct atomic substitution (AS) for
3.125\% and the virtual crystal approximation (VCA) for 2\%, 3.125\%
and 4\%. The energies are listed in eV per unit formula.}
\end{table}

We conclude the discussion on Si doping by comparing results obtained
by atomic substitution (AS) to the virtual crystal approximation (VCA).
In \tabref{Si_doping}, we list the energies of the orthorhombic,
tetragonal and cubic phases with respect to the monoclinic phase,
for 2\% and 4\% Si-doped cases computed by VCA, and the 3.125\% Si-doped
case computed by AS and VCA. We find that VCA is in agreement with
AS for the tetragonal phase, and gives an acceptable result for the
orthorhombic phase. We have investigated the disagreement in the cubic
phase by first inspecting the environment of the Si dopant in the
case of AS. The eightfold coordination of hafnium persists for the
silicon dopant. We have then relaxed the structure again after slightly
displacing one of the neighboring oxygens, which has resulted in the
transformation of the cell into a tetragonal cell, indicating that
the cubic phase is unstable toward silicon doping. Hence, for the
remainder of our study, we do not discuss the behavior of the cubic
phase, which is also not observed in hafnia-based thin films.

\subsubsection{Hf/Zr mixing}

Moving to the other most widely used dopant in hafnia thin films,
we turn to Zr. In \tabref{Zr_doping} we list the energies of the
orthorhombic, tetragonal and cubic phases with respect to the monoclinic
phase for bulk Hf$_{x}$Zr$_{1-x}$O$_{2}$, where $x=1.00,0.75,0.50,0.25,0.00$.
For each case we compare the AS and the VCA results. For AS computations,
we have used 4-unit-formula (12-atom) cells, and replaced 0 - 4 Hf
atoms in the cell with Zr. For the 50\% mixing case, where 2 Hf atoms
per cell are substituted by Zr, we compute the energies for all possible
2-atom substitutions in the cell. These differently chosen pairs of
atoms lead to relaxed energies within 0.02 eV of each other per unit
formula, and the lowest such energy is reported for each phase.

\begin{table}
\begin{centering}
\begin{tabular}{cccccc}
\toprule 
\addlinespace[0.2cm]
 & HfO$_{2}$ & $x=0.75$ & $x=0.50$ & $x=0.25$ & ZrO$_{2}$\tabularnewline\addlinespace[0.2cm]
\midrule 
\addlinespace[0.2cm]
ortho (AS) & 0.08 & 0.08 & 0.08 & 0.07 & 0.08\tabularnewline\addlinespace[0.2cm]
\midrule 
\addlinespace[0.2cm]
ortho (VCA) & 0.08 & 0.07 & 0.07 & 0.07 & 0.08\tabularnewline\addlinespace[0.2cm]
\midrule 
\addlinespace[0.2cm]
tetra (AS) & 0.16 & 0.14 & 0.14 & 0.14 & 0.12\tabularnewline\addlinespace[0.2cm]
\midrule 
\addlinespace[0.2cm]
tetra (VCA) & 0.16 & 0.14 & 0.13 & 0.12 & 0.12\tabularnewline\addlinespace[0.2cm]
\midrule 
\addlinespace[0.2cm]
cubic (AS) & 0.26 & 0.25 & 0.23 & 0.22 & 0.21\tabularnewline\addlinespace[0.2cm]
\midrule 
\addlinespace[0.2cm]
cubic (VCA) & 0.26 & 0.20 & 0.19 & 0.19 & 0.21\tabularnewline\addlinespace[0.2cm]
\bottomrule
\end{tabular}
\par\end{centering}
\caption[Energies of the orthorhombic, tetragonal and cubic phases with respect
to the monoclinic phase, for Hf$_{x}$Zr$_{1-x}$O$_{2}$ where $x=$1.00,
0.75, 0.50, 0.25 and 0.00.]{\label{tab:Zr_doping}Energies of the orthorhombic, tetragonal and
cubic phases with respect to the monoclinic phase, for Hf$_{x}$Zr$_{1-x}$O$_{2}$
where $x=$1.00, 0.75, 0.50, 0.25 and 0.00, as computed by atomic
substitution (AS) and virtual crystal approximation (VCA). The energies
are listed in eV per unit formula.}
\end{table}

We find that AS and VCA are in very good agreement for the orthorhombic
and tetragonal phases, as in the Si-doped case. We could not determine
the cause of the differences between these two methods for the cubic
case, but since the cubic phase does not appear to participate in
ferroelectricity of hafnia thin films, we have decided to leave this
question for future research.

\subsection{Effects of strain on doped HfO$_{2}$\label{subsec:Effects-of-strain}}

\subsubsection{Matching planes for bulk phases}

It has been observed that ferroelectric hafnia thin films have large
numbers of tetragonal grains in the early stages of growth \citep{tomida2006dielectric,boscke2009increasing,boscke2011ferroelectricity,muller2012ferroelectricity}.
This is understood to be caused by a reduction of the relative energy
of the tetragonal phase through its low surface energy \citep{garvie1978stabilization,batra2016stabilization}
and enhanced by doping. We have shown that for 3 - 4\% Si doping,
the tetragonal phase is significantly stabilized which agrees with
previous computational studies \citep{fischer2008stabilization,lee2008firstprinciples}.
We have also shown that for Zr mixing above 25\%, the tetragonal phase
is also stabilized, again in agreement with prior works \citep{lee2008firstprinciples,materlik2015theorigin}.
In \subsecref{Thin-film}, we will confirm that the interface energy
of the tetragonal phase with a common metal electrode is ind\textcolor{black}{eed
competitive} with the other phases. Hence it is reasonable to think
that a significant fraction of the initial grains during film growth
are tetragonal. Our hypothesis is that, after the deposition of the
top electrode and during thermal annealing, some portion of these
tetragonal grains transform into orthorhombic grains. We now investigate
this scenario in more detail and show that it is plausible for certain
doping ranges. Our main physical assumption will be that during the
potential transformation of a tetragonal grain into other phases,
the grain is geometrically\textit{ confined} within the film by the
surrounding grains: that it cannot change its in-plane area significantly
during the transformation.

\begin{figure}
\begin{centering}
\includegraphics[width=1\columnwidth]{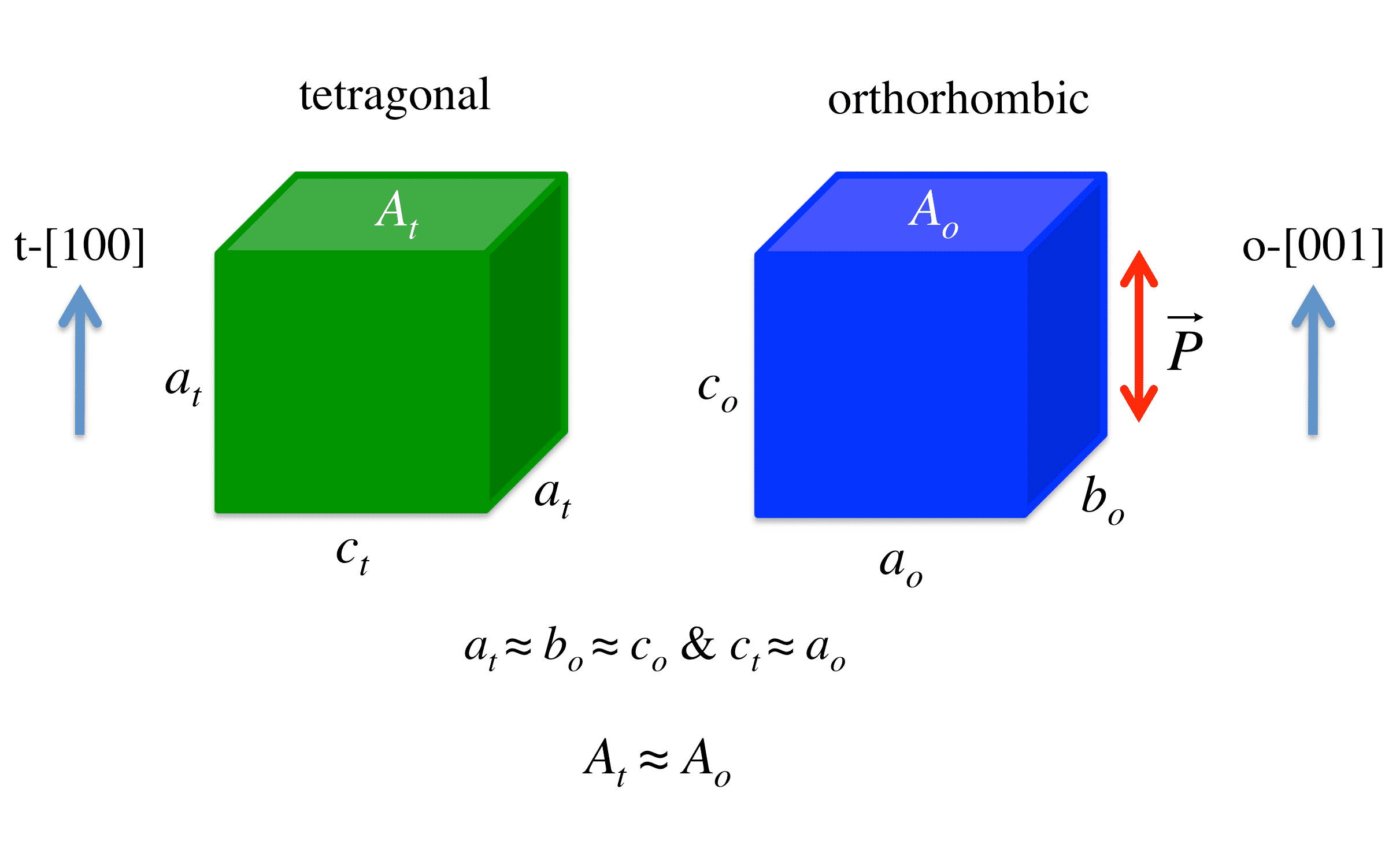}
\par\end{centering}
\caption[Conventional unit cells of the tetragonal and the orthorhombic phases,
and the approximate equalities between their lattice parameters and
lattice planes.]{\label{fig:Matchings}Conventional unit cells of the tetragonal and
the orthorhombic phases, and the approximate equalities between their
lattice parameters and lattice planes. The polarization vector lies
in the $\left[001\right]$ direction of the orthorhombic phase. A
$\left[100\right]$-oriented tetragonal grain can transform into a
$\left[001\right]$-oriented out-of-plane polarized orthorhombic grain
by a set of small changes in the lattice parameters. The quantities
$A_{t}$ and $A_{0}$ refer to the constrained planar areas of the
two phases.}
\end{figure}

In order for an out-of-plane polarized ($[001]$ oriented) orthorhombic
grain to form without a large change in the in-plane lattice parameters,
the parent tetragonal grain needs to have the orientation $\left[100\right]$
or $\left[010\right]$ (which are physically equivalent). We demonstrate
these matchings pictorially in \figref{Matchings}. The short sides
of the tetragonal phase ($a_{t}$) and the orthorhombic phase ($b_{o},c_{o}$)
are similar in length; and the long sides of these two phases ($c_{t}$
and $a_{o}$) are also similar in length (see \tabref{Lat_para} for
computed values). Therefore the tetragonal cell can transform into
the orthorhombic cell by slightly elongating $c_{t}$ and slightly
contracting one of the $a_{t}$.

Repeating the same analysis for the monoclinic phase, we eliminate
the $\text{m}\left[001\right]\longleftrightarrow\text{o}\left[001\right]$
transformation because of the mismatch in lengths, and the $\text{m}\left[010\right]\longleftrightarrow\text{o}\left[001\right]$
transformation because of the mismatch in the angles between the in-plane
lattice vectors. Therefore the constrained-area transformations that
can lead to a $[001]$ oriented orthorhombic phase are:

\[
\text{m}\left[100\right]\longleftrightarrow\text{t}\left[100\right]\longleftrightarrow\text{o}\left[001\right].
\]

\subsubsection{Effects of strain on undoped grains}

\begin{figure}
\begin{centering}
\includegraphics[width=0.75\columnwidth]{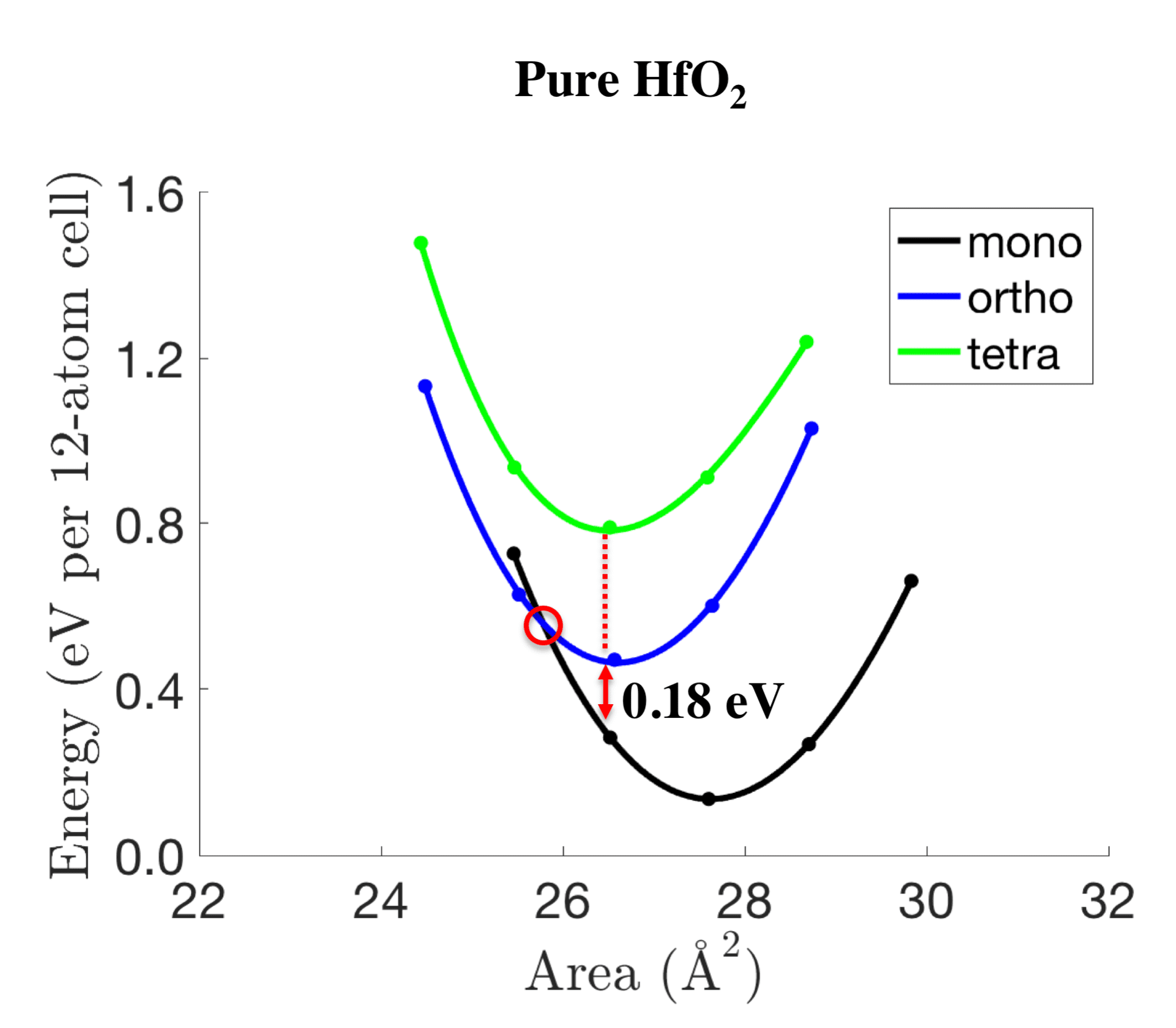}
\par\end{centering}
\caption[Energies of the monoclinic, orthorhombic and tetragonal phases vs
in-plane matching area for epitaxially strained bulk simulations of
pure HfO$_{2}$.]{\label{fig:Strain_HfO2}Energies of the monoclinic, orthorhombic
and tetragonal phases vs. in-plane matching area for epitaxially strained
bulk simulations of pure HfO$_{2}$. For each phase, five data points
at -4\%, -2\%, 0\%, 2\% and 4\% in-plane strain are simulated (circular
marks). The curves are obtained by fitting cubic polynomials to these
five data points. The energy difference between the orthorhombic and
the monoclinic phases at the optimized area of the $\text{t}\left[100\right]$
grain is equal to 0.18 eV per 12-atom cell, and labelled in the figure.
The zero of energy is chosen arbitrarily.}
\end{figure}

To investigate the likelihood of the $\text{tetra}\rightarrow\text{mono}$
and the $\text{tetra}\rightarrow\text{ortho}$ transformations, we
have simulated epitaxially strained phases of hafnia via computational
relaxations of bulk hafnia strained to pre-specified lattice parameters.
For each of the three phases, we have applied -4\%, -2\%, 0\%, 2\%
and 4\% biaxial strain \textcolor{black}{to each of the in-pla}ne
lattice parameters with respect to their unstrained values, and relaxed
the third lattice parameter as well as all the atomic positions. In
\figref{Strain_HfO2}, we plot the energies of the three phases of
HfO$_{2}$ versus the area of the matching plane. For each phase,
we fit a third degree polynomial to the five data points we have obtained
to generate a smooth curve.

A $\text{t}\left[100\right]$ grain with energy-optimal in-plane area
may transform into the orthorhombic and the monoclinic phases without
changing its area, which would be represented in \figref{Strain_HfO2}
as a downward jump from the bottom of the green curve to a point on
either the blue or the black curve. Because at the optimized area
of the $\text{t}\left[100\right]$ grain the monoclinic phase is 0.18
eV lower than the orthorhombic phase, the likelihood of the $\text{tetra}\rightarrow\text{mono}$
transformation should be much higher than the likelihood of the $\text{tetra}\rightarrow\text{ortho}$
transformation. We also circle in \figref{Strain_HfO2} the point
at which the curves that correspond to the monoclinic and the orthorhombic
phases cross. That point corresponds to a $3\%$ compressive biaxial
strain with respect to the $\text{t}\left[100\right]$ grain. Therefore,
in the absence of a mechanism that generates such a compressive strain,
the grain is expected to transform into a $\text{m}\left[100\right]$
grain during annealing.

To promote the transformation to the $\text{o}\left[001\right]$ instead,
Batra \textit{et al}. introduced an electric field \citep{batra2017factors}
and showed that the orthogonal phase can be made favorable with the
application of fields that are experimentally feasible. Below, we
explore the effects of a different physical factor, the dopant kind
and density, on the energy versus area curves in \figref{Strain_HfO2}.

\subsubsection{Effects of strain on Si:HfO$_{2}$}

\begin{figure*}
\begin{centering}
\includegraphics[width=0.75\textwidth]{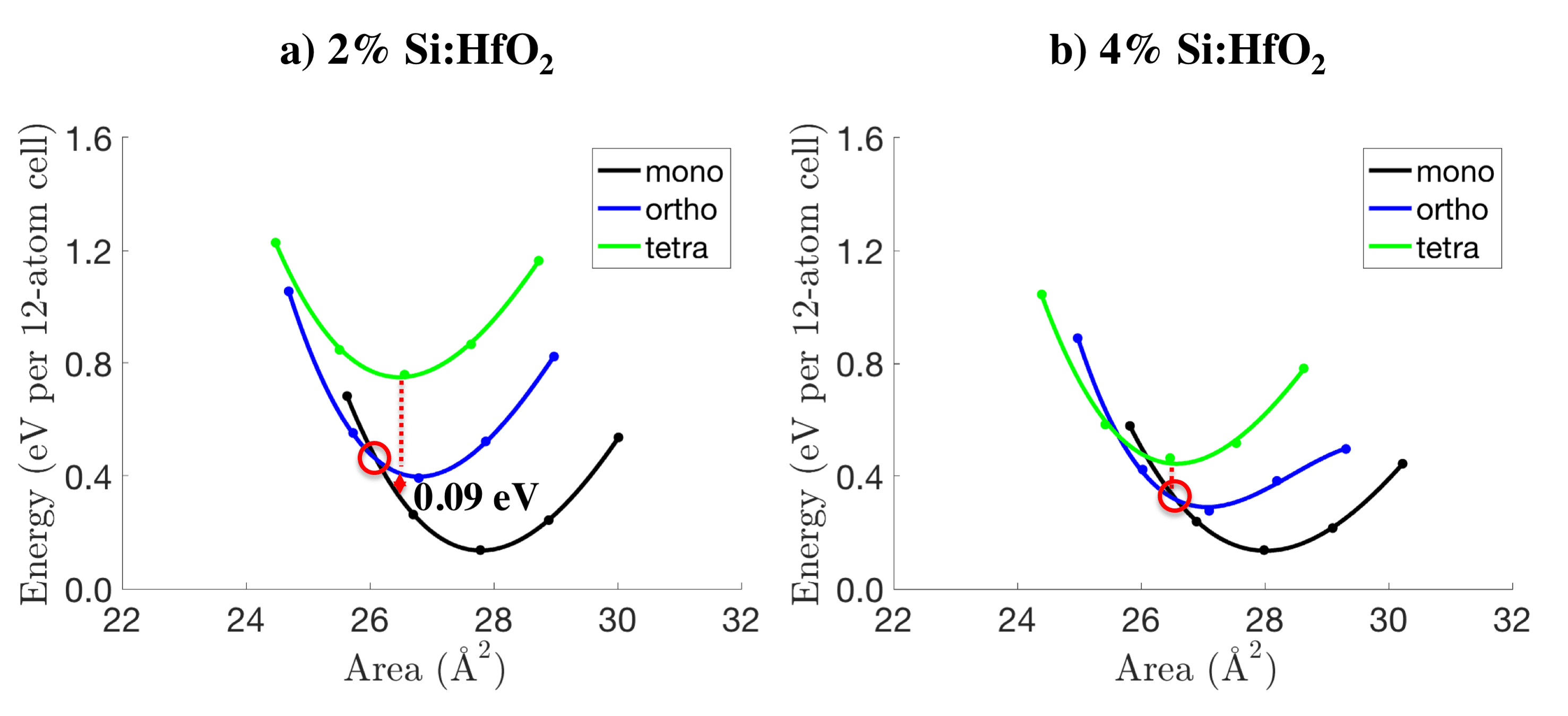}
\par\end{centering}
\caption[Energies of the monoclinic, orthorhombic and tetragonal phases vs
in-plane area for epitaxially strained bulk simulations of 2\% Si
doped and 4\% Si doped HfO$_{2}$.]{\label{fig:Strain_Si}Energies of the monoclinic, orthorhombic and
tetragonal phases vs in-plane matching area for epitaxially strained
bulk simulations of (a) 2\% Si doped and (b) 4\% Si doped HfO$_{2}$.
For each composition and phase, five data points at -4\%, -2\%, 0\%,
2\% and 4\% in-plane strain are simulated (circular marks). The curves
are obtained by fitting cubic polynomials to these five data points.
The energy difference between the orthorhombic and the monoclinic
phases at the optimized area of the $\text{t}\left[100\right]$ grain
is labelled in the figure in (a), and is equal to zero in (b). The
zero of energy is chosen arbitrarily.}
\end{figure*}

We repeat the same set of simulations for the 2\% and the 4\% Si doped
HfO$_{2}$ as modeled by the VCA. We present the results in \figref{Strain_Si}.

We observe in \figref{Strain_Si} (a) that for the 2\% Si doped HfO$_{2}$,
the energy difference between the orthorhombic and the monoclinic
phases at the optimized area of $\text{t}\left[100\right]$ grains
is 0.09 eV, which is lower than the undoped value of 0.18 eV. The
mono/ortho crossing occurs at 1\% compressive strain as opposed to
3\% in the undoped case. Hence the formation of the $\text{o}\left[001\right]$
grains through the $\text{t}\left[100\right]$ grains is favored by
Si doping. In the case of 4\% Si doping shown in \figref{Strain_Si}
(b), the mono/ortho crossing occurs at zero strain relative to the
optimal tetragonal in-plane area. Thus from a purely energetic point
of view, an optimized $\text{t}\left[100\right]$ grain has equal
chance of transforming into an $\text{o}\left[001\right]$ grain or
an $\text{m}\left[100\right]$ grain.

\begin{table}
\begin{centering}
\begin{tabular}{cccc}
\toprule 
\addlinespace[0.2cm]
 & $\ $HfO$_{2}$$\ $ & $\ $2\% Si$\ $ & $\ $4\% Si$\ $\tabularnewline\addlinespace[0.2cm]
\midrule 
\addlinespace[0.2cm]
\% strain where $E\left(\text{o}\right)=E\left(m\right)$ & -1.3\% & -0.7\% & 0.2\%\tabularnewline\addlinespace[0.2cm]
\midrule 
\addlinespace[0.2cm]
$E\left(\text{t}\right)-E\left(\text{o}\right)$ (eV) & 0.32 & 0.34 & 0.12\tabularnewline\addlinespace[0.2cm]
\midrule 
\addlinespace[0.2cm]
$E\left(\text{t}\right)-E\left(\text{m}\right)$ (eV) & 0.50 & 0.43 & 0.12\tabularnewline\addlinespace[0.2cm]
\bottomrule
\end{tabular}
\par\end{centering}
\caption[Summary of the results described in Figures \ref{fig:Strain_HfO2}
and \ref{fig:Strain_Si}.]{\label{tab:Strain_Si}Key numerical results described in Figures
\ref{fig:Strain_HfO2} and \ref{fig:Strain_Si}. The strain values
reported are biaxial strain with respect to the optimized in-plane
area of the $\left[100\right]$ oriented tetragonal phase when the
energies of the orthorhombic and the monoclinic phases coincide. The
energy differences are taken at the optimized in-plane area of the
$\left[100\right]$ oriented tetragonal phase and reported in eV per
12-atom cell.}
\end{table}

In \tabref{Strain_Si} we summarize our findings on the epitaxially
strained Si-doped HfO$_{2}$. As the doping concentration increases,
the energy difference between the orthorhombic and the monoclinic
phases at the optimized in-plane area of $\text{t}\left[100\right]$
grains decreases. At 4\% doping, the energies of the $\text{o}\left[001\right]$
and $\text{m}\left[100\right]$ grains coincide for the in-plane area
that is optimized for $\text{t}\left[100\right]$ grains. However,
the $\text{tetra}\rightarrow\text{mono}$ transformation that keeps
the area fixed increases the volume of the cell by 5\%, whereas the
$\text{tetra}\rightarrow\text{ortho}$ transformation that keeps the
area fixed decreases the volume by 1\%. Therefore, in the presence
of a top electrode that provides additional out-of-plane confinement,
the $\text{tetra}\rightarrow\text{ortho}$ transformation may have
a further advantage compared to the $\text{tetra}\rightarrow\text{mono}$
transformation. Our findings offer an explanation for the experimental
observation that 3-4\% Si doped films that are subjected to high temperature
annealing with a top electrode have ferroelectric properties; i.e.,
>2\% doping (in the case of silicon) and pre-annealing deposition
of the top electrode are necessary conditions for ferroelectricity
(see \secref{Experiment6}).

\subsubsection{Effects of strain on Hf$_{x}$Zr$_{1-x}$O$_{2}$}

\begin{figure*}
\begin{centering}
\includegraphics[width=0.75\textwidth]{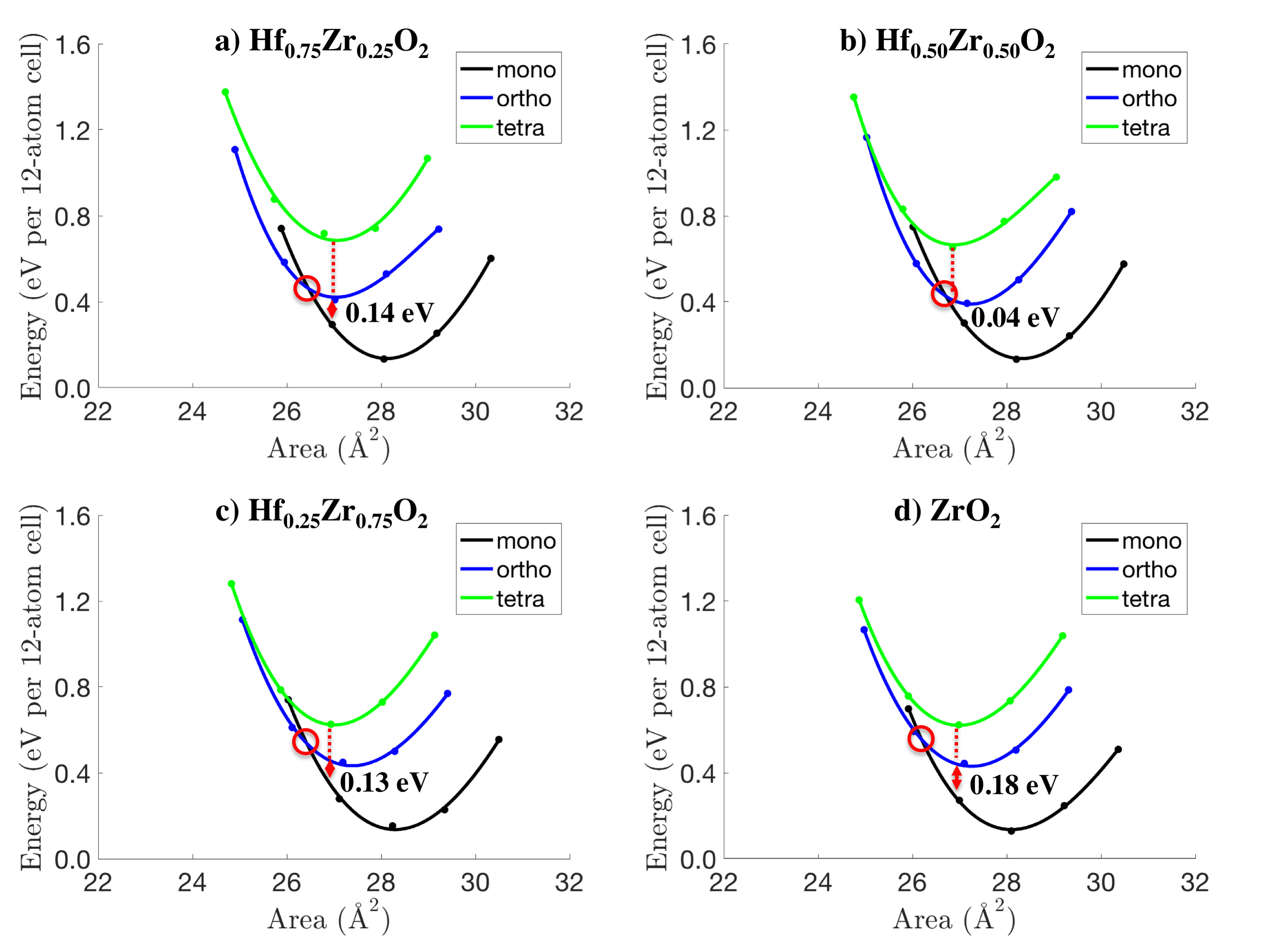}
\par\end{centering}
\caption[Energies of the monoclinic, orthorhombic and tetragonal phases vs
in-plane area for epitaxially strained bulk simulations of Hf$_{0.75}$Zr$_{0.25}$O$_{2}$,
Hf$_{0.50}$Zr$_{0.50}$O$_{2}$, Hf$_{0.25}$Zr$_{0.75}$O$_{2}$
and pure ZrO$_{2}$.]{\label{fig:Strain_Zr}Energies of the monoclinic, orthorhombic and
tetragonal phases vs in-plane matching area for epitaxially strained
bulk simulations of (a) Hf$_{0.75}$Zr$_{0.25}$O$_{2}$, (b) Hf$_{0.50}$Zr$_{0.50}$O$_{2}$,
(c) Hf$_{0.25}$Zr$_{0.75}$O$_{2}$ and (d) pure ZrO$_{2}$. For
each composition and phase, five data points at -4\%, -2\%, 0\%, 2\%
and 4\% in-plane strain are simulated (circular marks). The curves
are obtained by fitting cubic polynomials to these five data points.
The energy difference between the orthorhombic and the monoclinic
phases at the optimized area of the $\text{t}\left[100\right]$ grain
is labelled in the figure in for each case. The zero of energy is
chosen arbitrarily.}
\end{figure*}

As one of the most common hafnia derivatives that has successfully
been used as a ferroelectric thin film, we repeat the above analysis
of strain effects for Hf$_{x}$Zr$_{1-x}$O$_{2}$. We present our
results in \figref{Strain_Zr}. We find that the energy difference
between the orthorhombic and the monoclinic phases at the optimized
area of $\text{t}\left[100\right]$ grains is (a) 0.14 eV for $x=0.75$,
(b) 0.04 eV for $x=0.50$, (c) 0.13 eV for $x=0.25$ and (d) 0.18
eV for pure ZrO$_{2}$ (per 12-atom cell). Hence HfZrO$_{4}$ ($x=\nicefrac{1}{2}$)
presents the most suitable situation for the $\text{tetra}\rightarrow\text{ortho}$
transformation.

In \tabref{Strain_Zr} we report our relevant results for Hf$_{x}$Zr$_{1-x}$O$_{2}$.
The energy difference between the orthorhombic and the monoclinic
phases at the optimized in-plane area of $\text{t}\left[100\right]$
grains is minimized at $x=0.50$. For this case, the energies of the
$\text{o}\left[001\right]$ and $\text{m}\left[100\right]$ grains
coincide for the in-plane area that is 1\% compressively strained
with respect to the optimized area for $\text{t}\left[100\right]$
grains. Without the strain, the $\text{tetra}\rightarrow\text{mono}$
transformation is preferred to the $\text{tetra}\rightarrow\text{ortho}$
transformation by 0.04 eV per 12-atom cell. However, the former increases
the volume by 3\%, whereas the latter increases the volume by 1\%.
Therefore the confinement effects provided by the top electrode may
favor the $\text{tetra}\rightarrow\text{ortho}$ transformation over
the $\text{tetra}\rightarrow\text{mono}$ transformation. As in the
case of Si doping, our findings explain the experimental observation
that 30 - 60\% Zr doped films that are annealed with a capping electrode
present ferroelectric properties (see \secref{Experiment6}).

\begin{table}
\begin{centering}
\begin{tabular}{>{\centering}m{0.25\columnwidth}ccccc}
\toprule 
\addlinespace[0.2cm]
 & $\ $HfO$_{2}$$\ $ & $x=0.75$ & $x=0.50$ & $x=0.25$ & $\ $ZrO$_{2}$$\ $\tabularnewline\addlinespace[0.2cm]
\midrule 
\addlinespace[0.2cm]
\% strain where

$E\left(\text{o}\right)=E\left(m\right)$ & -1.3\% & -1.0\% & -0.4\% & -1.0\% & -1.5\%\tabularnewline\addlinespace[0.2cm]
\midrule 
\addlinespace[0.2cm]
$E\left(\text{t}\right)-E\left(\text{o}\right)$ (eV) & 0.32 & 0.27 & 0.26 & 0.17 & 0.18\tabularnewline\addlinespace[0.2cm]
\midrule 
\addlinespace[0.2cm]
$E\left(\text{t}\right)-E\left(\text{m}\right)$ (eV) & 0.50 & 0.43 & 0.30 & 0.30 & 0.36\tabularnewline\addlinespace[0.2cm]
\bottomrule
\end{tabular}
\par\end{centering}
\caption{\label{tab:Strain_Zr}Key numerical results described in Figures \ref{fig:Strain_HfO2}
and \ref{fig:Strain_Zr} for Hf$_{x}$Zr$_{1-x}$O$_{2}$ where $x=$1.00,
0.75, 0.50, 0.25 and 0.00. The strain values reported are biaxial
strain with respect to the optimized in-plane area of the $\left[100\right]$
oriented tetragonal phase when the energies of the orthorhombic and
the monoclinic phases coincide. The energy differences are taken at
the optimized in-plane area of the $\left[100\right]$ oriented tetragonal
phase, and reported in eV per 12-atom cell.}
\end{table}

In addition, we observe that for Hf$_{0.25}$Zr$_{0.75}$O$_{2}$
and ZrO$_{2}$, the energy versus strain curves that represent the
tetragonal and the orthorhombic phases lie closer to each other (see
\figref{Strain_Zr}). In \tabref{Strain_Zr}, we notice that the energy
difference between the optimized $\text{t}\left[100\right]$ grains
and the $\text{o}\left[100\right]$ grains with the same area is lowest
for high Zr:Hf ratios. Because the orthorhombic Pca2$_{1}$ space
group is a subgroup of the tetragonal P4$_{2}$/nmc space group, the
proximity in their energies promotes antiferroelectricity \citep{reyes-lillo2014antiferroelectricity}.
This supports the experimental observation of antiferroelectric behavior
in thin films with higher Zr content (see \secref{Experiment6}).

\subsubsection{Effects of strain on (Al, Ge, Ti, La):HfO$_{2}$}

To conclude this section, we repeat the strain analysis for four additional
dopants: Al, Ge, Ti and La. Our results are summarized in \tabref{Strain_other},
and the energy versus in-plane matching area plots are in the Supplementary
Material.

We observe that the effect of strain on the Al-doped bulk HfO$_{2}$
is very similar to that on the Si-doped bulk HfO$_{2}$. The strain
values at which the energies of the orthorhombic and the monoclinic
phases coincide for these two dopants at 4\% doping are close to zero
(0.2\% for Si and 0.3\% for Al); therefore, the transformation from
the tetragonal phase to the orthorhombic phase during annealing is
expected to be robust for both dopants. This is in agreement with
experiments which have found that the ferroelectric properties of
Si- and Al-doped HfO$_{2}$ are similar \citep{schroeder2014impactof,park2017effectof}.

We also observe that the 2\% and 4\% doping percentages for Ge, Ti
and La do not improve on the required strain values for the crossing
of energies of the orthorhombic and the monoclinic phases compared
to undoped HfO$_{2}$. Therefore, we predict that Ge and Ti may not
perform as well as Si and Al as dopants in HfO$_{2}$ in terms of
promoting ferroelectricity at low doping concentrations even though
their atomic radii are very close to those of Si and Al. Our similar
prediction regarding La is at odds with experiments to date in which
La:HfO$_{2}$ has yielded high $P_{r}$ values \citep{schroeder2014impactof,starschich2017anextensive}.
However, in these experiments, the doping concentration was greater
than $5\%$, which is beyond the reliable range of our VCA method
for non-isovalent elements. We leave a detailed analysis of the case
of La for future research.

\begin{table*}
\begin{centering}
\begin{tabular}{cccccccccc}
\toprule 
\addlinespace[0.2cm]
 & $\ $HfO$_{2}$$\ $ & $\ $2\% Al$\ $ & $\ $4\% Al$\ $ & $\ $2\% Ge$\ $ & $\ $4\% Ge$\ $ & $\ $2\% Ti$\ $ & $\ $4\% Ti$\ $ & $\ $2\% La$\ $ & $\ $4\% La$\ $\tabularnewline\addlinespace[0.2cm]
\midrule 
\addlinespace[0.2cm]
\% strain where $E\left(\text{o}\right)=E\left(m\right)$ & -1.3\% & -0.4\% & 0.3\% & -1.6\% & -2.0\% & -1.3\% & -1.6\% & -1.5\% & -1.4\%\tabularnewline\addlinespace[0.2cm]
\midrule 
\addlinespace[0.2cm]
$E\left(\text{t}\right)-E\left(\text{o}\right)$ (eV) & 0.32 & 0.29 & 0.07 & 0.32 & 0.34 & 0.28 & 0.29 & 0.31 & 0.30\tabularnewline\addlinespace[0.2cm]
\midrule 
\addlinespace[0.2cm]
$E\left(\text{t}\right)-E\left(\text{m}\right)$ (eV) & 0.50 & 0.35 & 0.07 & 0.58 & 0.65 & 0.44 & 0.52 & 0.51 & 0.50\tabularnewline\addlinespace[0.2cm]
\bottomrule
\end{tabular}
\par\end{centering}
\caption{\label{tab:Strain_other} Summary of the energy vs. strain results
for (Al, Ge, Ti, La):HfO$_{2}$. The strain values reported are biaxial
strain with respect to the optimized in-plane area of the $\left[100\right]$
oriented tetragonal phase when the energies of the orthorhombic and
the monoclinic phases coincide. The energy differences are taken at
the optimized in-plane area of the $\left[100\right]$ oriented tetragonal
phase, and reported in eV per 12-atom cell. Energy vs. in-plane matching
area curves for these four dopants and two doping percentages are
presented in the Supplementary Material.}
\end{table*}

Our analysis in this section supports our hypothesis that some of
the initially formed tetragonal grains transform into out-of-plane
polarized orthorhombic grains during thermal annealing. This requires
these grains to be confined in-plane by the surrounding grains, and
out-of-plane by the bottom and top electrodes. The ideal doping range
for this transformation is $\sim4\%$ for Si and Al, and $\sim50\%$
for Zr.

\subsection{Thin film simulations\label{subsec:Thin-film}}

In addition to the combined effects of doping and strain, we have
performed investigations on interface effects. In hafnia-based thin
films, ferroelectricity occurs when the film is $\sim$8-24 nm thick,
and the grains are generally a few nm in size. This makes finite-size
effects potentially important. Surfaces of ZrO$_{2}$ and HfO$_{2}$
have been studied experimentally \citep{pitcher2005energycrossovers,chevalier2009thetetragonalmonoclinic}
and theoretically \citep{garvie1978stabilization,mukhopadhyay2006firstprinciples,luo2008combined,luo2009monoclinic}
prior to the discovery of ferroelectricity in these films, with a
focus on the monoclinic and tetragonal phases. A recent study has
included the polar orthorhombic Pca2$_{1}$ phase into a first principles
investigation of surfaces of hafnia \citep{batra2016stabilization}.

\begin{figure}
\begin{centering}
\includegraphics[width=0.6\columnwidth]{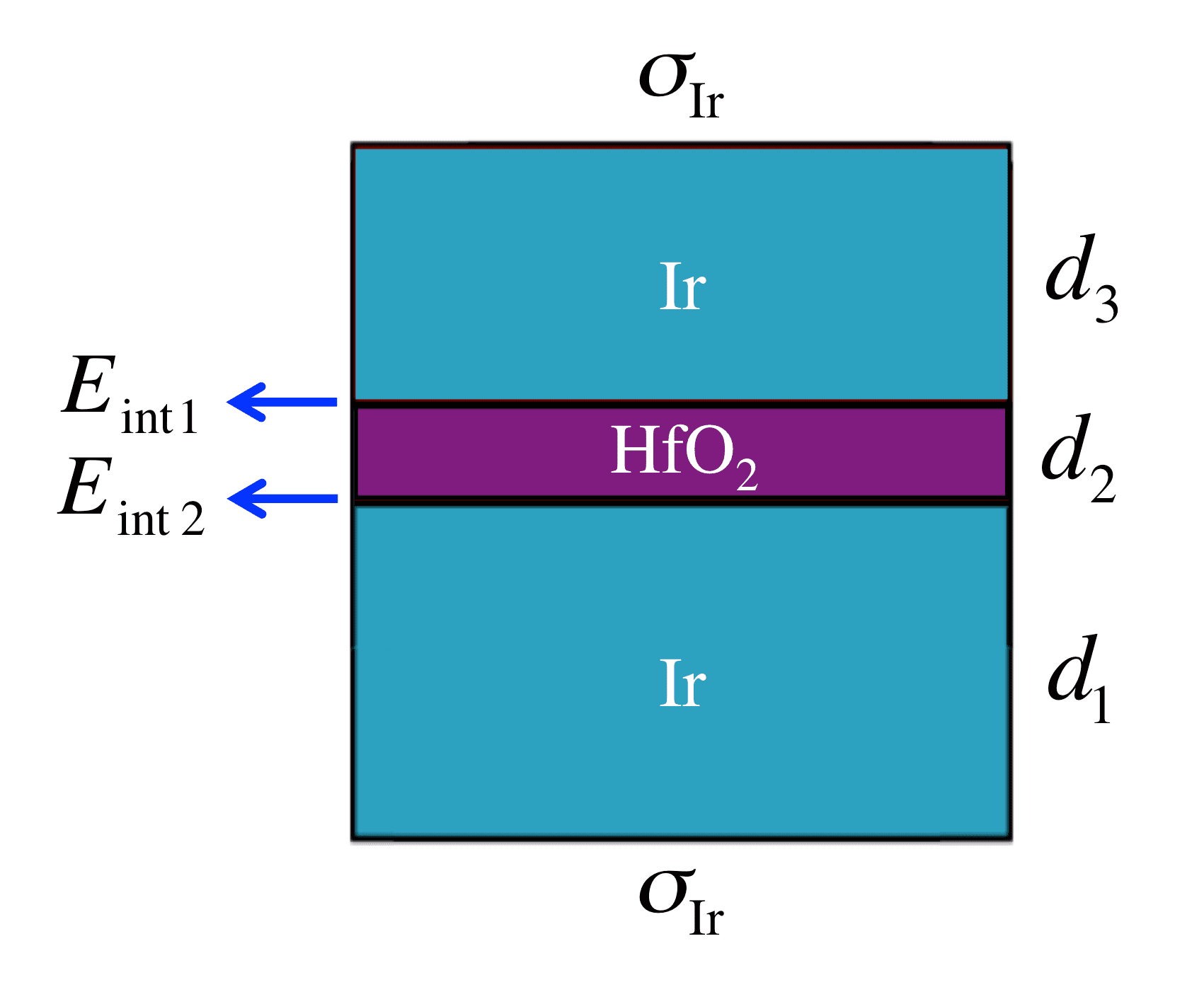}
\par\end{centering}
\caption[Schematic demonstration of a Ir/HfO$_{2}$/Ir stack simulation.]{\label{fig:Stack}Schematic demonstration of a Ir/HfO$_{2}$/Ir stack
simulation. $d_{1}$, $d_{2}$ and $d_{3}$ denote the thicknesses
of the components of the stack, $\sigma_{\text{Ir}}$ denotes the
surface energy of Ir, and $E_{\text{int 1}}$ and $E_{\text{int 2}}$
denote interface energies of the two Ir/HfO$_{2}$ interfaces.}
\end{figure}

Our goal is to compute the energies of the interfaces between relevant
phases of hafnia and typical electrodes such as TiN and Ir. In \figref{Stack},
we schematically depict an Ir/HfO$_{2}$/Ir stack. In order to isolate
thin film effects from strain effects, we fix the in-plane lattice
parameters to the lattice parameters of the HfO$_{2}$ phase in the
orientation that we choose to study. We have found that the lattice
parameters of HfO$_{2}$ are in the range of 5.04-5.30 $\text{\AA}$
(see \tabref{Lat_para}). On the other hand, typical electrodes used
with hafnia thin films, e.g. TiN and Ir, have lattice constants of
4.24 $\text{\AA}$ and 3.90 $\text{\AA}$, respectively. To faithfully
use these electrode lattice constants, we would need to simulate very
large supercells to create heterostructures where no significant strain
occurs on either the metal or the oxide. However, we believe that
such a calculation is not needed as a first pass, since epitaxial
growth is not actually observed in the experimental systems. Therefore,
we study the interfaces using much more reasonably sized $\sqrt{2}\times\sqrt{2}$
cells of Ir ($a=5.52$ $\text{\AA}$) with HfO$_{2}$, where the Ir
is strained to match various phases and orientations of HfO$_{2}$.
A similar TiN cell would have a lattice constant of $6.00$ $\text{\AA}$
and thus require a huge compressive strain, so we drop TiN from this
initial study and focus on Ir. Lastly, we emphasize that these are
\textit{model} calculations: the theoretical simulation has periodic
boundary conditions and thus is always epitaxial while the experimental
interfacial structure is much more complex, non-epitaxial, and unknown
with any precision at present. Moreover, in our calculations, the
Ir layers are significantly strained, and thus they represent only
an idealized model representation of a metal electrode rather than
the actual material used in the experiment. Our aim is to use a first-principles
approach to the interfacial energetics to gauge their \textit{approximate}
size and \textit{possible} importance under the assumption that the
interaction of HfO$_{2}$ with this theoretical model electrode is
a good proxy to the actual interfacial interaction. Although Ir is
employed less frequently than TiN in the experiments, it fits better
into our approach because it undergoes small relaxations when interfaced
with HfO$_{2}$. In contrast, our trial simulations of TiN/HfO$_{2}$
interfaces exhibited large distortions and formed cross-interface
chemical bonds between Ti and O atoms.

In order to compute the interfacial energies in a stack as shown in
\figref{Stack}, we first define the surface energy of the free-standing
Ir thin film as 
\begin{equation}
2\sigma_{\text{Ir}}=E_{\text{film}}^{\left(\text{Ir}\right)}\left(d\right)-d\times E_{\text{bulk}}^{\left(\text{Ir}\right)},\label{eq:surf}
\end{equation}
where $E_{\text{film}}^{\left(\text{Ir}\right)}\left(d\right)$ is
the computed energy of the free-standing strained Ir film, $E_{\text{bulk}}^{\left(\text{Ir}\right)}$
is the energy per u.c. of the bulk strained Ir, and $d$ is the thickness
of the Ir film in units of unit cells. An accurate way to extract
$\sigma_{\text{Ir}}$ is to compute $E_{\text{film}}^{\left(\text{Ir}\right)}\left(d\right)$
as a function of $d$ and fit a straight line, treating $\sigma_{\text{Ir}}$
and $E_{\text{bulk}}^{\left(\text{Ir}\right)}$ as fitting parameters
\citep{fiorentini1996extracting}. For this task, we have used $d=2,3,4,5,6\text{ u.c.}$
(each unit cell of Ir consists of two atomic layers). The resulting
surface energies are listed in \tabref{Int_energies}.

Once we have found $\sigma_{\text{Ir}}$ for a given phase and orientation
of HfO$_{2}$, we calculate the interfacial energies $E_{int}$ as
\begin{equation}
E_{\text{int 1}}+E_{\text{int 2}}=E_{\text{stack}}-\left(d_{1}+d_{3}\right)E_{\text{bulk}}^{\left(\text{Ir}\right)}-d_{2}E_{\text{bulk}}^{\left(\text{HfO}_{2}\right)}-2\sigma_{\text{Ir}},\label{eq:int}
\end{equation}
where $E_{\text{stack}}$ is the computed energy of the final materials
stack, $E_{\text{bulk}}^{\left(\text{Ir}\right)}$ is the energy per
u.c. of the bulk strained Ir, $E_{\text{bulk}}^{\left(\text{HfO}_{2}\right)}$
is the energy per u.c. of bulk HfO$_{2}$, and the thicknesses $d_{1}$,
$d_{2}$ and $d_{3}$ are shown in \figref{Stack} and are in unit
cells. We note that we can only compute the sum of the two interfacial
energies using this approach; if the interfaces are physically identical,
then a single interface energy becomes available.

The relaxed configuration of the Ir/HfO$_{2}$/Ir stack with $d_{1}=d_{2}=d_{3}=2\text{ u.c.}$
where HfO$_{2}$ is in the monoclinic-$\left[001\right]$ configurati\textcolor{black}{on
is shown in \figref{Simcell}. }To find the lowest energy interfaces
for a given phase and orientation, we have first simulated possible
surface terminations of HfO$_{2}$. In order to include an integer
number of unit cells in the thin film (i.e., stoichiometric hafnia),
we have restricted the terminations to be Hf on one end and O on the
other (Hf-$\ldots$-OO) or O terminated on both ends (O-$\ldots$-O).
Prior work on zirconia has shown that the energy of an O-$\ldots$-O
terminated slab is lower in energy than a Zr-$\ldots$-OO terminated
slab by 13.0 eV per in-plane cell for $\text{t}\left[001\right]$
films of free-standing ZrO$_{2}$ \citep{eichler2004firstprinciples}.
Analogous results have been reported for hafnia as well \citep{batra2016stabilization}.
We have found a very similar (and huge) value of 13.4 eV for free-standing
HfO$_{2}$ by simulating 2 u.c. thick $\text{t}\left[001\right]$
films. To check that the Hf-$\ldots$-OO termination remains high-energy
for oxide/metal interfaces, we have simulated HfO$_{2}$/Ir stacks
with the cubic phase for (Ir)-Hf-$\ldots$-OO-(Ir), (Ir)-OO-$\ldots$-Hf-(Ir)
and (Ir)-O-$\ldots$-O-(Ir) terminations. We have found that the (Ir)-Hf-$\ldots$-OO-(Ir)
and (Ir)-OO-$\ldots$-Hf-(Ir) stacks are 6.3 and 7.8 eV per in-plane
cell higher in energy than the (Ir)-O-$\ldots$-O-(Ir) stack, respectively.
Therefore, we have decided to restrict our studies to the O-$\ldots$-O
terminated HfO$_{2}$ films for all phases of hafnia. In order to
find the lowest energy interface for each HfO$_{2}$ phase and orientation,
we have run relaxations for the top and bottom interfaces separately,
using a $2\times2$ lateral grid of initial HfO$_{2}$ positions relative
to Ir for each case. After finding the optimal coordinates for the
top and bottom interfaces separately, we have joined them to make
the Ir/HfO$_{2}$/Ir stacks, and then fully relaxed the atomic positions
(except for the surface u.c. of Ir).

\begin{table}
\begin{centering}
\begin{tabular}{>{\centering}m{0.5\columnwidth}cc}
\toprule 
\addlinespace[0.2cm]
 & 2$\sigma_{\text{Ir}}$ (eV) & $E_{\text{int 1}}+E_{\text{int 2}}$ (eV)\tabularnewline\addlinespace[0.2cm]
\midrule 
\addlinespace[0.2cm]
monoclinic-$\left[001\right]$ & 8.8 & 5.7\tabularnewline\addlinespace[0.2cm]
\midrule 
\addlinespace[0.2cm]
monoclinic-$\left[100\right]$ & 9.5 & 9.8\tabularnewline\addlinespace[0.2cm]
\midrule 
\addlinespace[0.2cm]
orthorhombic-$\left[001\right]$ & 8.6 & 7.4\tabularnewline\addlinespace[0.2cm]
\midrule 
\addlinespace[0.2cm]
orthorhombic-$\left[100\right]$ & 7.7 & 7.0\tabularnewline\addlinespace[0.2cm]
\midrule 
\addlinespace[0.2cm]
orthorhombic-$\left[010\right]$ & 8.8 & 13.4\tabularnewline\addlinespace[0.2cm]
\midrule 
\addlinespace[0.2cm]
tetragonal-$\left[001\right]$ & 7.8 & 13.3\tabularnewline\addlinespace[0.2cm]
\midrule 
\addlinespace[0.2cm]
tetragonal-$\left[100\right]$ & 8.6 & 6.9\tabularnewline\addlinespace[0.2cm]
\midrule 
\addlinespace[0.2cm]
cubic-$\left[001\right]$ & 7.8 & 11.4\tabularnewline\addlinespace[0.2cm]
\bottomrule
\end{tabular}
\par\end{centering}
\caption[Interface energies for Ir/HfO$_{2}$/Ir stacks for each HfO$_{2}$
phase and orientation.]{\label{tab:Int_energies}Surface energies for strained iridium slabs
and interface energies for Ir/HfO$_{2}$/Ir stacks for each HfO$_{2}$
phase and orientation computed via \eqref{surf} and \eqref{int},
respectively. Energies are listed in eV per in-plane cell.}
\end{table}

We have studied the interfaces of Ir and pure HfO$_{2}$ in the monoclinic,
tetragonal, orthorhombic and cubic phases, in all possible inequivalent
principle orientations. The only exception is the $\text{m}\left[010\right]$
orientation, which is excluded because of the non-orthogonal in-plane
lattice vectors. To extract $E_{\text{int 1}}+E_{\text{int 2}}$ accurately
from \eqref{int}, we have computed $E_{\text{stack}}\left(d_{1},d_{2},d_{3}\right)$
with $d_{1}=d_{3}=2,3,4\text{ u.c.}$ and $d_{2}=2,3,4\text{ u.c.}$,
fitting\textcolor{black}{{} a linear equation wi}th $E_{\text{bulk}}^{\left(\text{Ir}\right)}$,
$E_{\text{bulk}}^{\left(\text{HfO}_{2}\right)}$ and $E_{\text{int 1}}+E_{\text{int 2}}+2\sigma_{\text{Ir}}$
as the fitting parameters. After extracting $E_{\text{int 1}}+E_{\text{int 2}}+2\sigma_{\text{Ir}}$,
we have used the $\sigma_{\text{Ir}}$ values found earlier to compute
$E_{\text{int 1}}+E_{\text{int 2}}$ for each case. We have found
that all phases and orientations are mechanically stable in thin film
form, with modest relaxations at both Ir/HfO$_{2}$ interfaces. We
list the interface energies we have found in \tabref{Int_energies}.
Our results are in the range of 5-13 eV, which is comparable to the
4-12 eV range found for (twice) the surface energies of HfO$_{2}$
\citep{batra2016stabilization}.

\begin{figure}
\begin{centering}
\includegraphics[width=1\columnwidth]{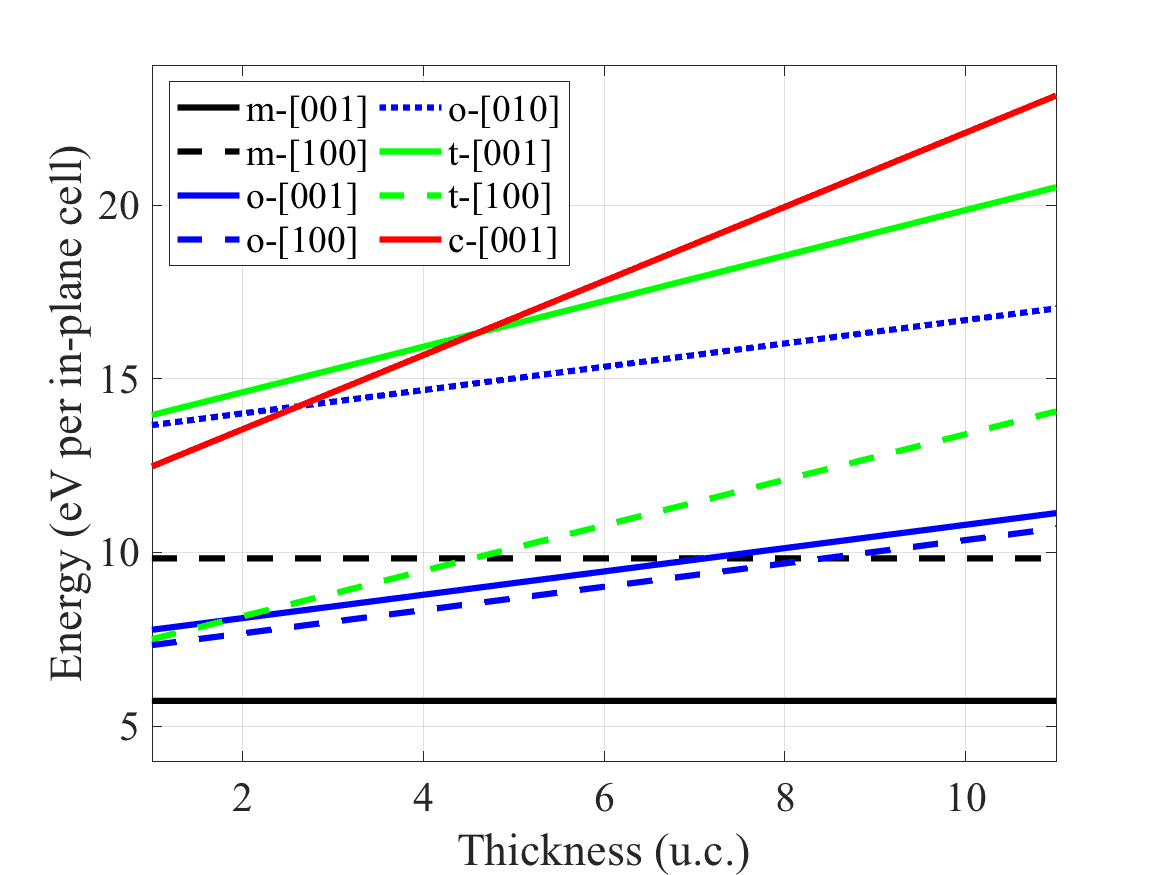}
\par\end{centering}
\caption[Thin film energies computed via the model described in as a function
of film thickness for Ir/HfO$_{2}$/Ir stacks.]{\label{fig:FilmE}Thin film energies computed via the model described
in \eqref{film} as a function of film thickness for Ir/HfO$_{2}$/Ir
stacks. The zero of the bulk energies is taken as the monoclinic phase,
causing the lines corresponding to this phase (black) to be flat.
The orthorhombic, tetragonal and cubic phases are represented by blue,
green and red lines, respectively.}
\end{figure}

With the interface energies we have obtained via \eqref{int} and
listed in \tabref{Int_energies}, we build an energetic model for
variable-thickness films based on our interfacial energies and bulk
hafnia energies using the formula:
\begin{equation}
E_{\text{film}}^{\left(\text{ph-or}\right)}\left(d\right)=E_{\text{int 1}}^{\left(\text{ph-or}\right)}+E_{\text{int 2}}^{\left(\text{ph-or}\right)}+dE_{\text{bulk}}^{\left(\text{ph}\right)},\label{eq:film}
\end{equation}
where $E_{\text{film}}^{\left(\text{ph-or}\right)}\left(d\right)$
is the energy of the thin film of HfO$_{2}$ in a given phase and
orientation ``ph-or'', $d$ is the thickness of the hafnia in u.c.,
and $E_{\text{bulk}}^{\left(\text{ph}\right)}$ is the energy of one
unit cell of HfO$_{2}$ in phase ``ph''.

We summarize the results of this model in \figref{FilmE} as a plot
of energy vs. thickness for each of the phase-orientation pairs we
have studied. We define $E_{\text{bulk}}^{\left(\text{mono}\right)}$
to be zero, so the lines that correspond to the monoclinic phase are
horizontal. We observe that, because of variations in the interface
energies that are of the order of a few eV per in-plane cell, several
crossings occur. Even though the $\text{m}\left[001\right]$ grains
remain as the lowest energy configuration for all thicknesses, we
find that for ultra-thin films, $\text{o}\left[001\right]$, $\text{o}\left[100\right]$
and $\text{t}\left[100\right]$ grains are also competitive (see also
\tabref{Int_energies} for the interface energy values). If some $\text{t}\left[100\right]$
grains are formed initially, they may become kinetically trapped as
the film grows further. The energy of the $\text{t}\left[100\right]$
film crosses the energies of the $\text{o}\left[001\right]$ films
approximately at a thickness of 2 unit cells. For thicker films, $\text{o}\left[001\right]$
grains are more favorable than $\text{t}\left[100\right]$ grains,
but less favorable than $\text{m}\left[100\right]$. Hence, this simple
model predicts that $\text{t}\left[100\right]$ grains may initially
form during growth and then transform into $\text{o}\left[001\right]$,
$\text{o}\left[100\right]$, $\text{m}\left[001\right]$ or $\text{m}\left[100\right]$
grains during annealing. The matching plane arguments from \subsecref{Effects-of-strain}
disfavor the transformation into $\text{o}\left[100\right]$, $\text{m}\left[001\right]$;
hence we would expect $\text{t}\left[100\right]\rightarrow\text{m}\left[100\right]$
and $\text{t}\left[100\right]\rightarrow\text{o}\left[001\right]$
transformations to dominate. Our thin film results thus agree with
the $\text{t}\left[100\right]\rightarrow\left(\text{\text{\text{m}\ensuremath{\left[100\right]},\text{o}\ensuremath{\left[001\right]}}}\right)$
picture above; however, the doping-induced modifications in the energy
vs. strain curves, presented in \subsecref{Effects-of-strain}, as
well as volumetric confinement, are necessary to favor the $\text{t}\left[100\right]\rightarrow\text{o}\left[001\right]$
transformation over the $\text{t}\left[100\right]\rightarrow\text{m}\left[100\right]$
transformation. In summary, our interfacial model indicates that interface
effects (1) are \textit{significant} and can \textit{dominate} in
the early stages of growth, and (2) do not explain the observed properties
of the films unless they are considered \textit{in conjunction} with
doping and strain effects. We expect these overall conclusions to
be true for all electrodes employed in the experiments including TiN.

Our model has a number of limitations. Two that can be addressed relatively
easily in future studies are: (1) The simulation for every phase and
orientation is done at the unstrained lattice parameters of that phase
and orientation. The electrode is assumed to be unaffected by strain
in any significant way. This assumption can be tested by applying
small strain to each phase and orientation and re-computing the interface
energies to check that they do not change in irregular ways. (2) The
films are assumed to stay exactly stoichiometric, i.e. an exact monolayer
(ML) of oxygen (and then a ML of hafnium) at both interfaces. Further
simulations can be run with 0, 0.5, 1.5 and 2 ML of oxygen at the
interfaces, yielding differing energy vs. thickness lines in \figref{FilmE}.

\section{Conclusion\label{sec:Conclusion6}}

We have conducted a first-principles study of doped hafnia with the
goal of understanding some of the experimental observations concerning
ferroelectricity from a structural point of view. We have described
the effects of various dopants on the energetics of bulk phases of
HfO$_{2}$. We have discussed in detail the structural changes that
are caused by Si doping. We have compared two computational methods
for modeling doping: atomic substitution (AS) and the virtual crystal
approximation (VCA). We have found that VCA compares well with AS
and used VCA to simulate the effects of epitaxial strain on doped
HfO$_{2}$. We have found that among 0\%, 2\% and \%4 Si doping, 4\%
doping provides the best conditions for initial $\text{tetragonal}\left[100\right]$
grains to transform into $\text{orthorhombic}\left[001\right]$ grains.
We have also found that for Hf$_{x}$Zr$_{1-x}$O$_{2}$, $x=0.5$
provides the most favorable conditions for the tetragonal $\left[100\right]$
$\rightarrow$ orthorhombic $\left[001\right]$ transformation. However,
for this transformation to be preferred over the tetragonal $\left[100\right]$
$\rightarrow$ monoclinic $\left[100\right]$ transformation, some
confinement needs to be present. In experiments, this confinement
is provided by a top electrode (typically TiN). Our findings provide
an explanation for common experimental observations for the optimal
ranges of doping in Si:HfO$_{2}$ and Hf$_{x}$Zr$_{1-x}$O$_{2}$.
Repeating the same analysis for Al, Ti, Ge and La, we have found that
Al:HfO$_{2}$ behaves similarly to Si:HfO$_{2}$; whereas Ti, Ge or
La doping only slightly modifies the strain response of pure HfO$_{2}$.
Finally, we have described a model to estimate the interface effects
for thin films of hafnia, based on \emph{ab initio} simulations of
Ir/HfO$_{2}$/Ir stacks. Our results offer interesting clues for how
the interface effects may be in play for the stabilization of the
ferroelectric phase in these films.

\section{Acknowledgements}

This work was supported by the National Science Foundation under Award
1609162 and by the grant MRSEC DMR-1119826. We thank the Yale Center
for Research Computing for guidance and use of the research computing
infrastructure, with special thanks to Stephen Weston and Andrew Sherman.
Additional computational support was provided by NSF XSEDE resources
via Grant TG-MCA08X007.

\bibliographystyle{apsrev}
\bibliography{Citations}

\end{document}


\chapter*{Supplementary Material for ``Causes of ferroelectricity in HfO$_{2}$
thin films''}
\begin{center}
{\large{}Mehmet Dogan$^{1,2,3,4}$, Nanbo Gong$^{1,5}$, Tso-Ping
Ma$^{1,5}$ and Sohrab Ismail-Beigi$^{1,2,6,7}$}\\
\par\end{center}

\begin{center}
$^{1}$Center for Research on Interface Structures and Phenomena,
Yale University, New Haven, Connecticut 06520, USA
\par\end{center}

\begin{center}
$^{2}$Department of Physics, Yale University, New Haven, Connecticut
06520, USA
\par\end{center}

\begin{center}
$^{3}$Department of Physics, University of California, Berkeley,
California 94720, USA
\par\end{center}

\begin{center}
$^{4}$Materials Science Division, Lawrence Berkeley National Laboratory,
Berkeley, California 94720, USA
\par\end{center}

\begin{center}
$^{5}$Department of Electrical Engineering, Yale University, New
Haven, Connecticut 06520, USA
\par\end{center}

\begin{center}
$^{6}$Department of Applied Physics, Yale University, New Haven,
Connecticut 06520, USA
\par\end{center}

\begin{center}
$^{7}$Department of Mechanical Engineering and Materials Science,
Yale University, New Haven, Connecticut 06520, USA
\par\end{center}

\section*{Oxygen coordination for dopants in HfO$_{2}$}

We list the dopant-O distances for the nearest oxygens in the 3.125\%
doped HfO$_{2}$ for C, N, N{*} (N -1e), Al, Al{*} (Al +1e), Ti and
Ge doping in \tabref{CN_oxygen_1}; and for Sr, Sr{*} (Sr +2e), Y,
Y{*} (Y +1e), La and La{*} (La +1e) doping in \tabref{CN_oxygen_2}.

\begin{table}[H]
\begin{centering}
\begin{tabular}{|c|c|c|c|c|c|c|c|c|c|}
\hline 
\noalign{\vskip0.2cm}
Phase & \multicolumn{8}{c|}{Nearest O neighbor distances ($\mathring{\text{A}}$)} & C. N.\tabularnewline[0.2cm]
\hline 
\noalign{\vskip0.2cm}
mono C & 1.28 & 1.31 & 1.36 & 2.67 & 2.70 & 2.74 & 2.85 &  & 3\tabularnewline[0.2cm]
\hline 
\noalign{\vskip0.2cm}
ortho C & 1.27 & 1.35 & 1.37 & 2.48 & 2.57 & 2.66 & 2.70 &  & 3\tabularnewline[0.2cm]
\hline 
\noalign{\vskip0.2cm}
tetra C & 1.45 & 1.45 & 1.45 & 1.45 & 2.86 & 2.86 & 2.86 & 2.86 & 4\tabularnewline[0.2cm]
\hline 
\noalign{\vskip0.2cm}
mono N & 1.26 & 1.31 & 1.71 & 2.62 & 2.79 & 2.85 & 2.96 &  & 2\tabularnewline[0.2cm]
\hline 
\noalign{\vskip0.2cm}
ortho N & 1.15 & 2.08 & 2.15 & 2,40 & 2.56 & 2.56 & 2.63 &  & 1\tabularnewline[0.2cm]
\hline 
\noalign{\vskip0.2cm}
tetra N & 1.24 & 1.24 & 1.95 & 2.18 & 2.74 & 2.74 & 2.75 & 2.76 & 2\tabularnewline[0.2cm]
\hline 
\noalign{\vskip0.2cm}
mono N{*} & 1.24 & 1.25 & 1.30 & 2.60 & 3.02 & 3.08 & 3.17 &  & 3\tabularnewline[0.2cm]
\hline 
\noalign{\vskip0.2cm}
ortho N{*} & 1.24 & 1.29 & 1.30 & 2.63 & 2.66 & 2.72 & 2.91 &  & 3\tabularnewline[0.2cm]
\hline 
\noalign{\vskip0.2cm}
tetra N{*} & 1.27 & 1.28 & 1.28 & 2.33 & 2.62 & 2.84 & 2.84 & 2.88 & 3\tabularnewline[0.2cm]
\hline 
\noalign{\vskip0.2cm}
mono Al & 1.87 & 1.92 & 1.95 & 2.03 & 2.06 & 2.16 & 2.81 &  & 5\tabularnewline[0.2cm]
\hline 
\noalign{\vskip0.2cm}
ortho Al & 1.91 & 1.96 & 1.97 & 2.00 & 2.01 & 2.09 & 2.79 &  & 5\tabularnewline[0.2cm]
\hline 
\noalign{\vskip0.2cm}
tetra Al & 1.83 & 1.83 & 1.83 & 1.83 & 2.64 & 2.64 & 2.64 & 2.64 & 4\tabularnewline[0.2cm]
\hline 
\noalign{\vskip0.2cm}
mono Al{*} & 1.88 & 1.94 & 1.95 & 2.04 & 2.06 & 2.19 & 2.82 &  & 5\tabularnewline[0.2cm]
\hline 
\noalign{\vskip0.2cm}
ortho Al{*} & 1.94 & 1.95 & 1.97 & 1.99 & 2.03 & 2.12 & 2.87 &  & 5\tabularnewline[0.2cm]
\hline 
\noalign{\vskip0.2cm}
tetra Al{*} & 1.83 & 1.83 & 1.83 & 1.83 & 2.71 & 2.71 & 2.71 & 2.71 & 4\tabularnewline[0.2cm]
\hline 
\noalign{\vskip0.2cm}
mono Ti & 1.92 & 1.96 & 2.06 & 2.07 & 2.10 & 2.25 & 2.27 &  & 5\tabularnewline[0.2cm]
\hline 
\noalign{\vskip0.2cm}
ortho Ti & 1.88 & 2.03 & 2.04 & 2.07 & 2.09 & 2.24 & 2.30 &  & 5\tabularnewline[0.2cm]
\hline 
\noalign{\vskip0.2cm}
tetra Ti & 1.88 & 1.88 & 1.88 & 1.88 & 2.58 & 2.58 & 2.58 & 2.58 & 4\tabularnewline[0.2cm]
\hline 
\noalign{\vskip0.2cm}
mono Ge & 1.88 & 1.88 & 1.94 & 1.95 & 1.99 & 2.22 & 2.84 &  & 5\tabularnewline[0.2cm]
\hline 
\noalign{\vskip0.2cm}
ortho Ge & 1.87 & 1.92 & 1.93 & 1.93 & 1.99 & 2.17 & 2.90 &  & 5\tabularnewline[0.2cm]
\hline 
\noalign{\vskip0.2cm}
tetra Ge & 1.81 & 1.81 & 1.81 & 1.81 & 2.69 & 2.69 & 2.69 & 2.69 & 4\tabularnewline[0.2cm]
\hline 
\end{tabular}
\par\end{centering}
\caption{\label{tab:CN_oxygen_1}List of (C, N, N{*}, Al, Al{*}, Ti, Ge)-O
bond lengths for each of the monoclinic, orthorhombic and tetragonal
phases, for the 3.125\% doped HfO$_{2}$. The number of oxygen neighbors
to the dopant (coordination number) is reported in the rightmost column.
It is assumed that if the distance between the two atoms is not much
larger than the sum of their atomic radii, the two atoms are coordinated.}
\end{table}

\begin{table}[H]
\begin{centering}
\begin{tabular}{|c|c|c|c|c|c|c|c|c|c|}
\hline 
\noalign{\vskip0.2cm}
Phase & \multicolumn{8}{c|}{Nearest O neighbor distances ($\mathring{\text{A}}$)} & C. N.\tabularnewline[0.2cm]
\hline 
\noalign{\vskip0.2cm}
mono Sr & 2.31 & 2.31 & 2.33 & 2.43 & 2.43 & 2.43 & 2.46 &  & 7\tabularnewline[0.2cm]
\hline 
\noalign{\vskip0.2cm}
ortho Sr & 2.32 & 2.35 & 2.40 & 2.41 & 2.43 & 2.45 & 2.46 &  & 7\tabularnewline[0.2cm]
\hline 
\noalign{\vskip0.2cm}
tetra Sr & 2.38 & 2.38 & 2.38 & 2.38 & 2.50 & 2.50 & 2.50 & 2.50 & 8\tabularnewline[0.2cm]
\hline 
\noalign{\vskip0.2cm}
mono Sr{*} & 2.32 & 2.32 & 2.38 & 2.43 & 2.46 & 2.47 & 2.49 &  & 7\tabularnewline[0.2cm]
\hline 
\noalign{\vskip0.2cm}
ortho Sr{*} & 2.32 & 2.38 & 2.42 & 2.42 & 2.45 & 2.46 & 2.52 &  & 7\tabularnewline[0.2cm]
\hline 
\noalign{\vskip0.2cm}
tetra Sr{*} & 2.40 & 2.40 & 2.40 & 2.40 & 2.55 & 2.55 & 2.55 & 2.55 & 8\tabularnewline[0.2cm]
\hline 
\noalign{\vskip0.2cm}
mono Y & 2.17 & 2.19 & 2.27 & 2.27 & 2.29 & 2.34 & 2.36 &  & 7\tabularnewline[0.2cm]
\hline 
\noalign{\vskip0.2cm}
ortho Y & 2.18 & 2.24 & 2.27 & 2.27 & 2.29 & 2.33 & 2.36 &  & 7\tabularnewline[0.2cm]
\hline 
\noalign{\vskip0.2cm}
tetra Y & 2.24 & 2.24 & 2.24 & 2.24 & 2.42 & 2.42 & 2.42 & 2.42 & 8\tabularnewline[0.2cm]
\hline 
\noalign{\vskip0.2cm}
mono Y{*} & 2.17 & 2.19 & 2.27 & 2.28 & 2.30 & 2.35 & 2.37 &  & 7\tabularnewline[0.2cm]
\hline 
\noalign{\vskip0.2cm}
ortho Y{*} & 2.19 & 2.27 & 2.28 & 2.30 & 2.34 & 2.38 & 2.50 &  & 7\tabularnewline[0.2cm]
\hline 
\noalign{\vskip0.2cm}
tetra Y{*} & 2.24 & 2.24 & 2.24 & 2.24 & 2.44 & 2.44 & 2.44 & 2.44 & 8\tabularnewline[0.2cm]
\hline 
\noalign{\vskip0.2cm}
mono La & 2.23 & 2.25 & 2.34 & 2.35 & 2.36 & 2.40 & 2.44 &  & 7\tabularnewline[0.2cm]
\hline 
\noalign{\vskip0.2cm}
ortho La & 2.25 & 2.30 & 2.34 & 2.36 & 2.39 & 2.40 & 2.42 &  & 7\tabularnewline[0.2cm]
\hline 
\noalign{\vskip0.2cm}
tetra La & 2.31 & 2.31 & 2.31 & 2.31 & 2.47 & 2.47 & 2.47 & 2.47 & 8\tabularnewline[0.2cm]
\hline 
\noalign{\vskip0.2cm}
mono La{*} & 2.24 & 2.25 & 2.33 & 2.38 & 2.39 & 2.43 & 2.45 &  & 7\tabularnewline[0.2cm]
\hline 
\noalign{\vskip0.2cm}
ortho La{*} & 2.22 & 2.31 & 2.38 & 2.38 & 2.39 & 2.44 & 2.47 &  & 7\tabularnewline[0.2cm]
\hline 
\noalign{\vskip0.2cm}
tetra La{*} & 2.32 & 2.32 & 2.32 & 2.32 & 2.49 & 2.49 & 2.49 & 2.49 & 8\tabularnewline[0.2cm]
\hline 
\end{tabular}
\par\end{centering}
\caption{\label{tab:CN_oxygen_2}List of (Sr, Sr{*}, Y, Y{*}, La and La{*})-O
bond lengths for each of the monoclinic, orthorhombic and tetragonal
phases, for the 3.125\% doped HfO$_{2}$. The number of oxygen neighbors
to the dopant (coordination number) is reported in the rightmost column.
It is assumed that if the distance between the two atoms is not much
larger than the sum of their atomic radii, the two atoms are coordinated.}
\end{table}

\section*{Energy vs matching area curves for (Al, Ge, Ti, La):HfO$_{2}$}

\begin{figure}
\begin{centering}
\includegraphics[width=0.9\columnwidth]{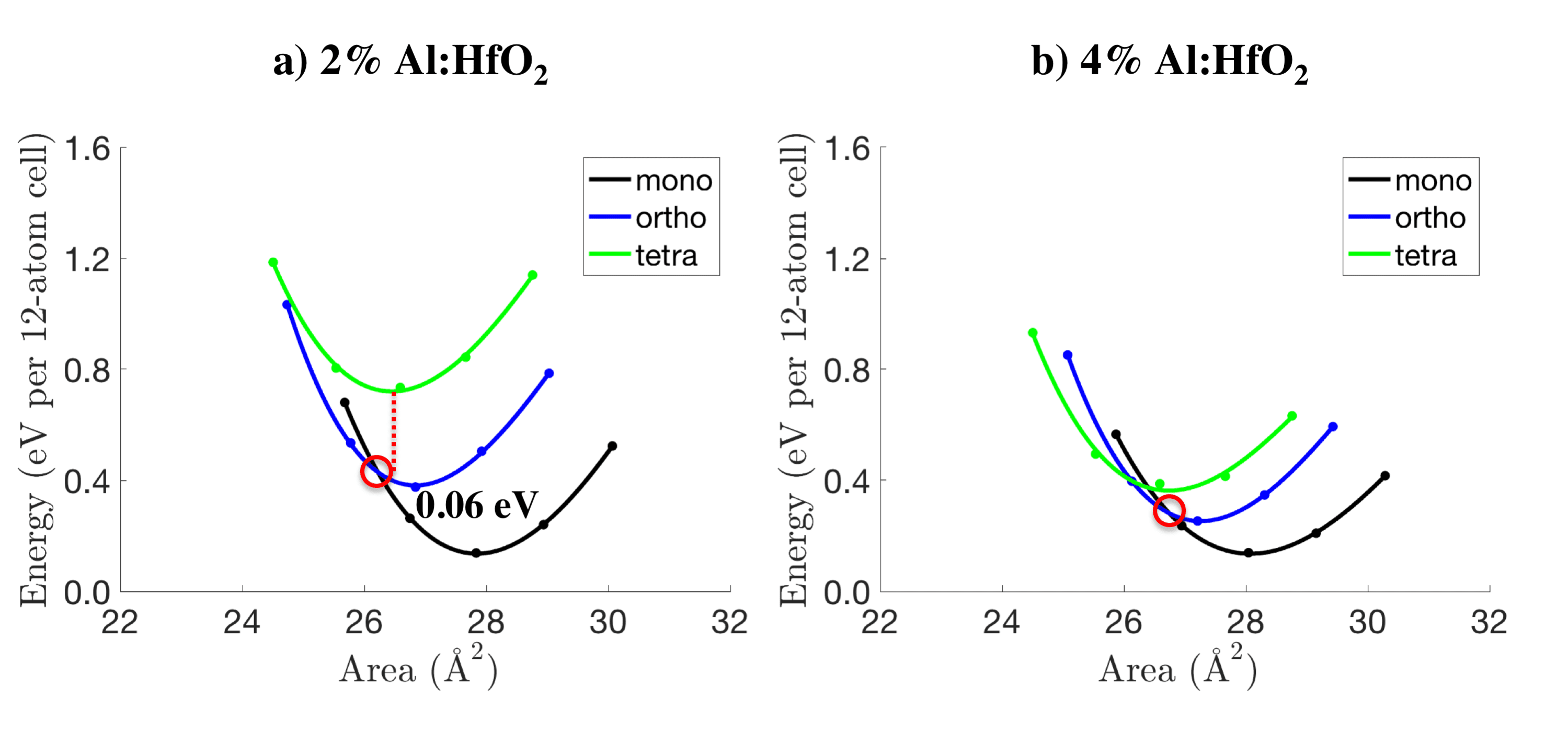}
\par\end{centering}
\caption{\label{fig:Strain_Al}Energies of the monoclinic, orthorhombic and
tetragonal phases vs in-plane matching area for epitaxially strained
bulk simulations of (a) 2\% Al doped and (b) 4\% Al doped HfO$_{2}$.
For each composition and phase, five data points at -4\%, -2\%, 0\%,
2\% and 4\% strain are chosen and computed (circular marks). The curves
are obtained by fitting cubic polynomials to these five data points.
The energy difference between the orthorhombic and the monoclinic
phases at the optimized area of the $\text{t}\left[100\right]$ grain
is labelled in the figure in (a), and is equal to zero in (b). The
zero of energy is chosen arbitrarily.}
\end{figure}

\begin{figure}
\begin{centering}
\includegraphics[width=0.9\columnwidth]{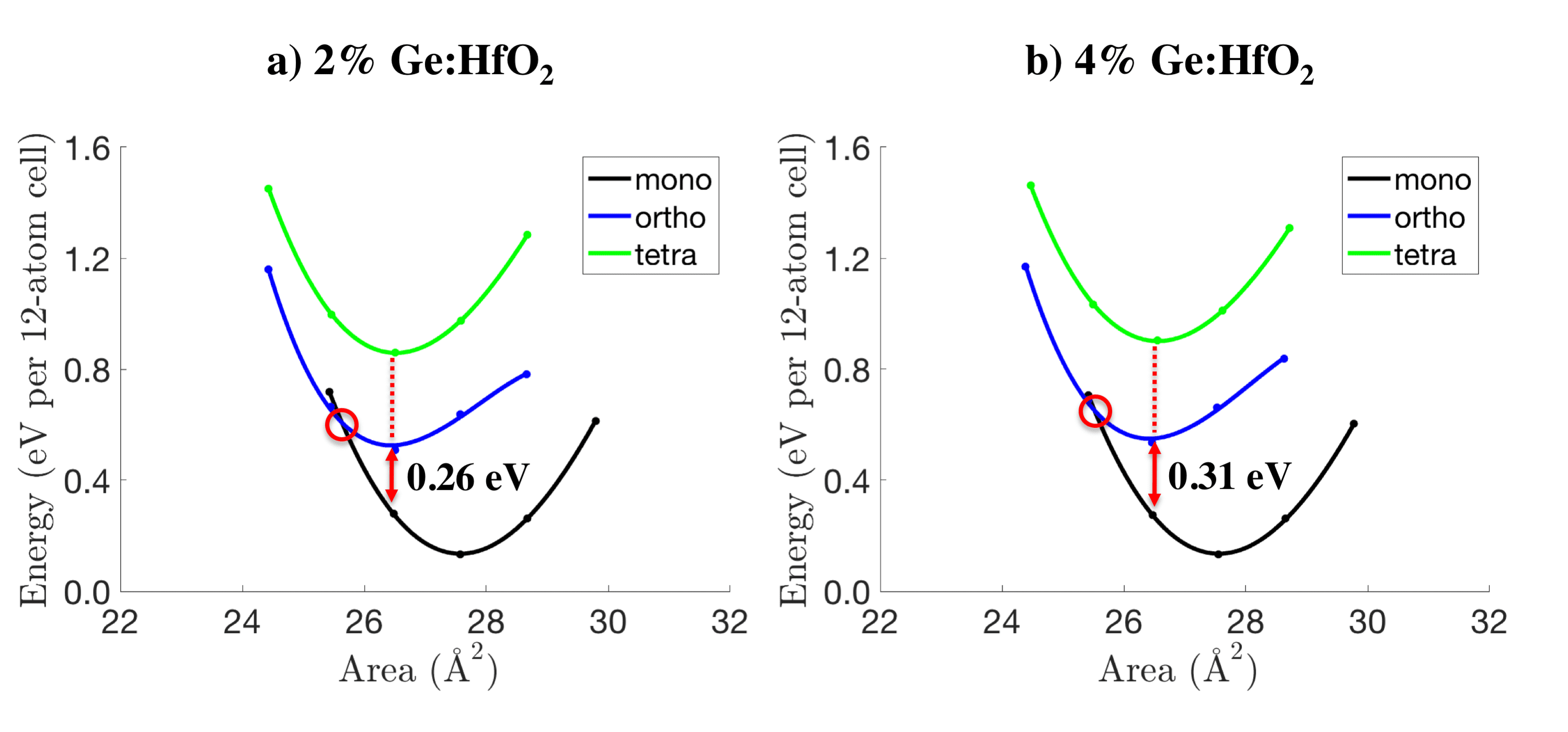}
\par\end{centering}
\caption{\label{fig:Strain_Ge}Energies of the monoclinic, orthorhombic and
tetragonal phases vs in-plane matching area for epitaxially strained
bulk simulations of (a) 2\% Ge doped and (b) 4\% Ge doped HfO$_{2}$.
For each composition and phase, five data points at -4\%, -2\%, 0\%,
2\% and 4\% strain are chosen and computed (circular marks). The curves
are obtained by fitting cubic polynomials to these five data points.
The energy difference between the orthorhombic and the monoclinic
phases at the optimized area of the $\text{t}\left[100\right]$ grain
is labelled in the figure. The zero of energy is chosen arbitrarily.}
\end{figure}

\begin{figure}
\begin{centering}
\includegraphics[width=0.9\columnwidth]{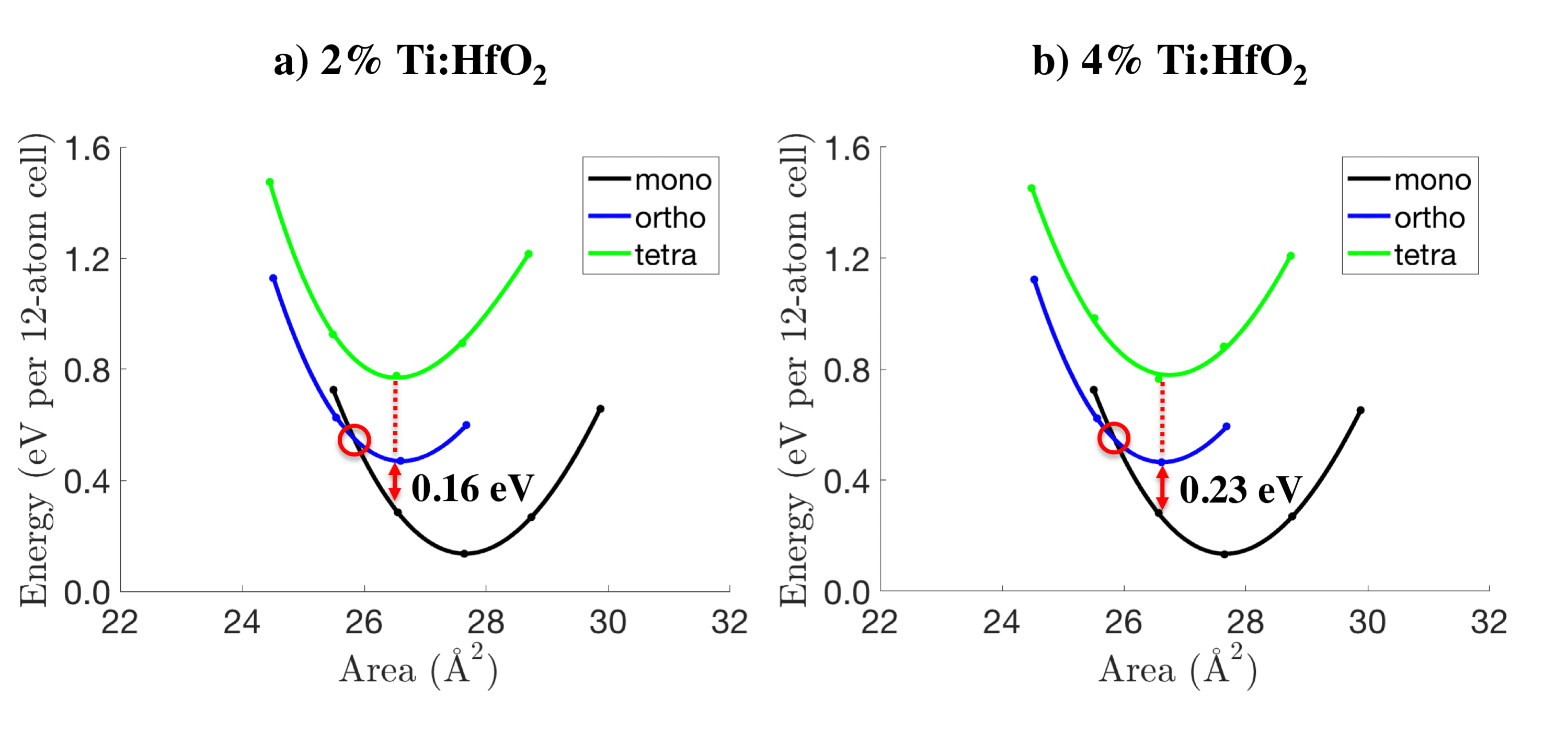}
\par\end{centering}
\caption{\label{fig:Strain_Ti}Energies of the monoclinic, orthorhombic and
tetragonal phases vs in-plane matching area for epitaxially strained
bulk simulations of (a) 2\% Ti doped and (b) 4\% Ti doped HfO$_{2}$.
For each composition and phase, five data points at -4\%, -2\%, 0\%,
2\% and 4\% strain are chosen and computed (circular marks). The curves
are obtained by fitting cubic polynomials to these five data points.
The energy difference between the orthorhombic and the monoclinic
phases at the optimized area of the $\text{t}\left[100\right]$ grain
is labelled in the figure. The zero of energy is chosen arbitrarily.}
\end{figure}

\begin{figure}
\begin{centering}
\includegraphics[width=0.9\columnwidth]{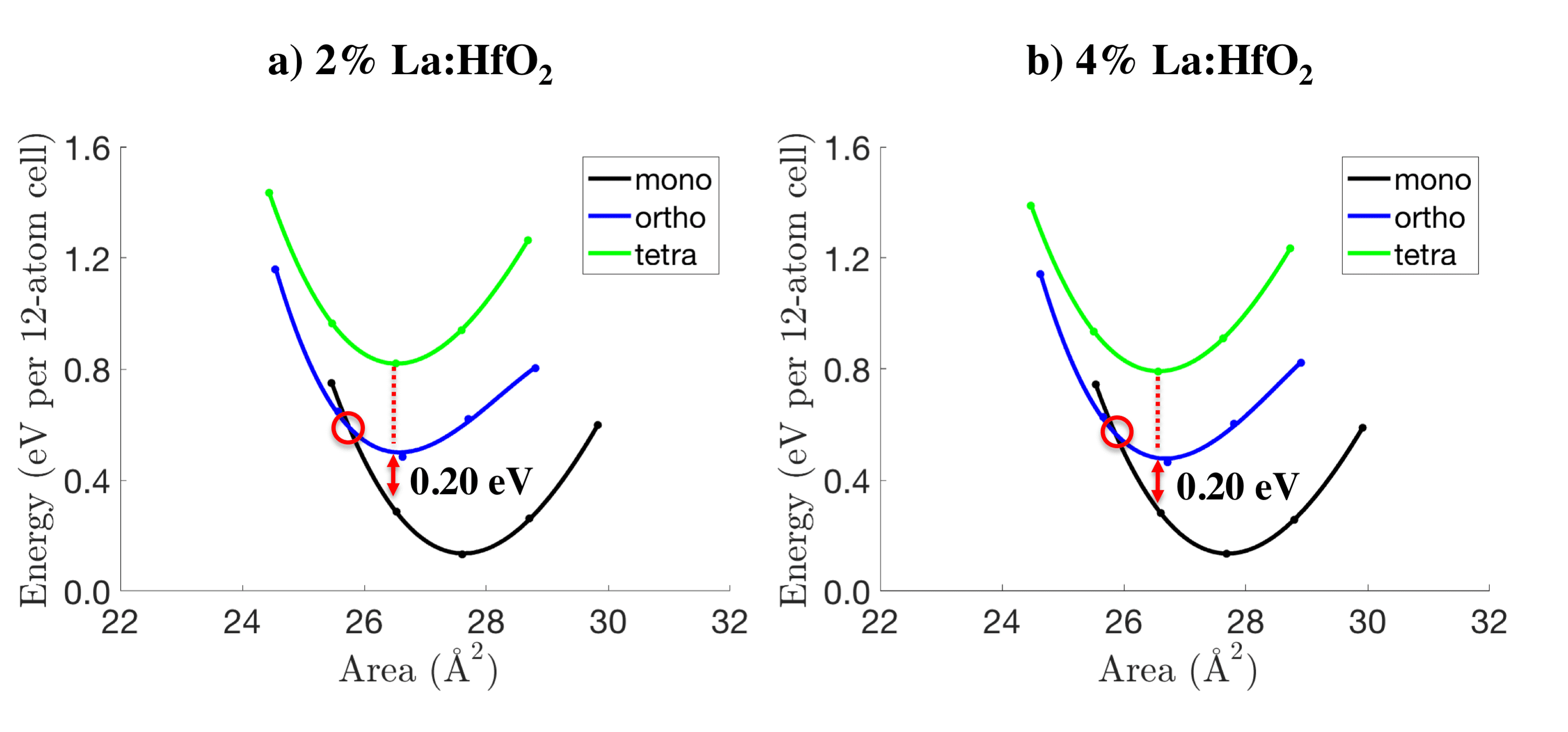}
\par\end{centering}
\caption{\label{fig:Strain_La}Energies of the monoclinic, orthorhombic and
tetragonal phases vs in-plane matching area for epitaxially strained
bulk simulations of (a) 2\% La doped and (b) 4\% La doped HfO$_{2}$.
For each composition and phase, five data points at -4\%, -2\%, 0\%,
2\% and 4\% strain are chosen and computed (circular marks). The curves
are obtained by fitting cubic polynomials to these five data points.
The energy difference between the orthorhombic and the monoclinic
phases at the optimized area of the $\text{t}\left[100\right]$ grain
is labelled in the figure. The zero of energy is chosen arbitrarily.}
\end{figure}